\documentclass[lettersize,journal]{IEEEtran}
\usepackage{amsmath,amsfonts}
\usepackage{algorithmic}
\usepackage{algorithm}
\usepackage{array}
\usepackage[caption=false,font=normalsize,labelfont=sf,textfont=sf]{subfig}
\usepackage{textcomp}
\usepackage{stfloats}
\usepackage{url}
\usepackage{verbatim}
\usepackage{graphicx}
\usepackage{cite}
\usepackage{xcolor}
\usepackage{textgreek}
\usepackage{hhline}
\usepackage{tabularx}
\usepackage{threeparttable}

\begin{document}

\title{A Second-Order Audio VCO-ADC with 103-dB-A Dynamic Range and Binary-Weighted Internal Architecture}

\author{Victor Medina,~\IEEEmembership{Graduate Student Member,~IEEE,} Ruben Garvi, Javier Granizo, Pedro Amaral and Luis Hernandez~\IEEEmembership{Senior Member,~IEEE}

\thanks{This research was funded by project PID2020-118804RB-I00 of the AEI, Spain.}%
\thanks{P.~Amaral is with Infineon Technologies Austria AG, 9500 Villach, Austria.}
\thanks{V.~Medina, R.~Garvi, J.~Granizo and L.~Hernandez are with the Department of Electronic Technology, Carlos III University, 28911  Leganes, Spain (email:luish@ing.uc3m.es).}}%

\markboth{}%
{Shell \MakeLowercase{\textit{et al.}}: A Sample Article Using IEEEtran.cls for IEEE Journals}

\maketitle

\begin{abstract}
One of the limitations of conventional VCO-ADCs is the restriction to first-order noise shaping. True-VCO architectures have been proposed to increase the noise-shaping order by cascading several VCO integrators, but without requiring analog feedback loops. A high noise-shaping order allows to reduce the input VCO frequency compared to a conventional VCO-ADC with similar dynamic range, which improves power consumption. Prior-art True-VCO architectures represent state variables either with a thermometer code or with a single-bit. Thermometer encoding is a natural choice when ring oscillators are selected as loop filter integrators. However, chip area restrictions force thermometer-encoded state variables to have few levels. A reduced number of levels in the state variables limits the dynamic range of True VCO-ADCs. In this paper, we show experimentally a second-order audio VCO-based ADC which uses ring oscillators as integrators but employs Gray and binary encoding for state variables. As a consequence, the complexity and area of the True-VCO architecture is reduced, breaking the barrier that limits the dynamic range of prior designs. The implemented chip shows a dynamic range of 103~dB achieving a peak SNDR of 76.5 dB-A with a power of 250 $\mu$W occupying 0.095 $\text{mm}^2$ in 130 nm CMOS.
\end{abstract}


\section{Introduction}
\begin{figure}[t]
	\centering
	\includegraphics[width=\columnwidth,keepaspectratio]{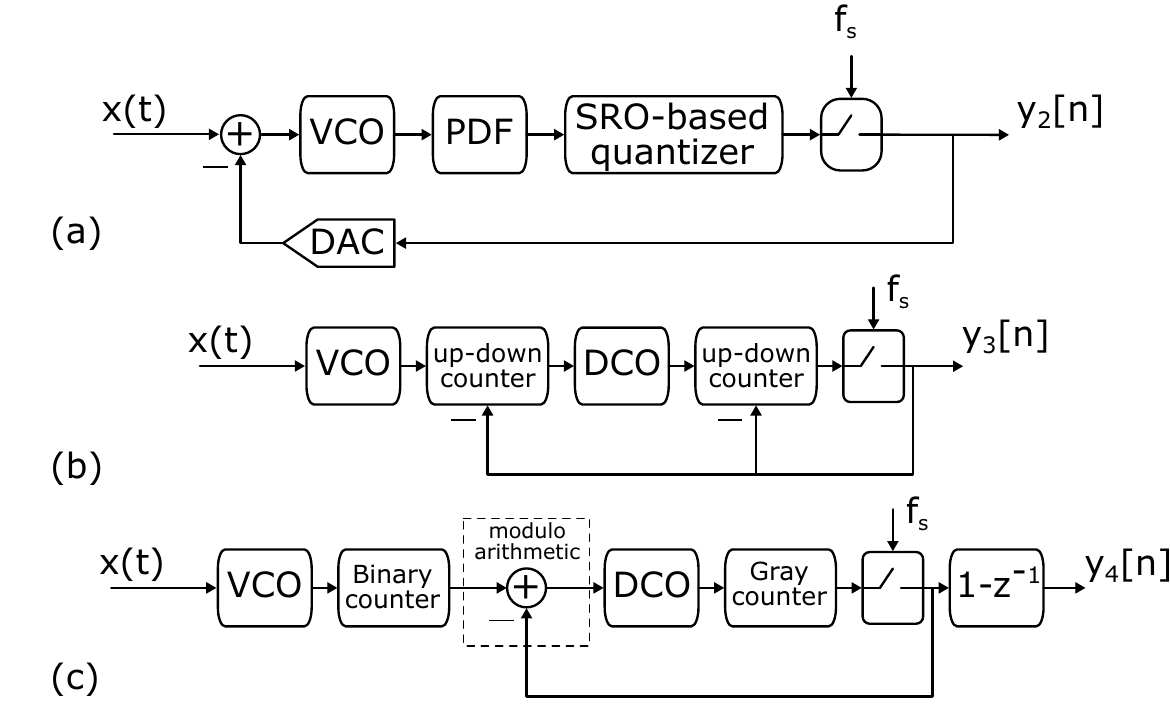}
	\caption{(a) Loop integrator approach, analog feedback DAC, (b) Up-down counter, (c) proposed binary weighted architecture.}
	\label{fig:comparison_prior_art}
\end{figure}

\IEEEPARstart{O}{versampled} converters based on Voltage Controlled Oscillators (VCO-ADCs) have become a popular architecture for sensors and biomedical signals \cite{whyandhow, rvco, carlosmls, cardestimestamp}. The advantages of VCO-ADCs rely on the small footprint of VCO circuitry and a mostly digital implementation \cite{colorines2}. VCO-ADCs employing a feedback DAC as in a Continuous Time Sigma Delta Modulator (CTSDM) have been proposed to compensate VCO nonlinearity and noise \cite{nan-sun-purely-vco-adc} (see Fig.~\ref{fig:comparison_prior_art} (a)). However, open-loop VCO-ADCs \cite{carlos_sensors, nus, galton, sensors_andres, jaewook_kim_tcasi } can also be a hardware efficient alternative when the connection to a high output impedance signal source is required. This is the typical case  of capacitive MEMS sensors  and measurements of bio-potentials originated in an electrode, such as ECG, EEG or neural probes. Open loop VCO-ADCs permit sharing the bias current of the VCO with an input preamplifier such as a transconductor or source follower, providing both high input impedance and low power. Although open loop architectures have been questioned due to the lack of linearity of Ring Oscillator (RO) VCOs, a number of techniques have been developed \cite{rombouts_electronic_letters,garvilin,bulkdlr} that mitigate VCO non-linearity with small power and area.  However, a problem common to many VCO-ADCs is the restriction to first-order noise shaping, especially when  the sampling rate is forced by industry standards (like in a MEMS microphone) or limited by technology (like in high speed data converters), \cite{optimization_borgmans}. In a first-order VCO-ADC, the oscillator effective frequency $f_e$, defined as the product of the RO rest frequency $f_0$ times the number of VCO phases $M$, is proportional to the resolution required \cite{colorines1}. Given that the VCO power consumption directly depends on the oscillation frequency, it may happen that the power consumed by the ADC is dictated by quantization noise instead of VCO phase noise \cite{cardes_noise}, which is a sub-optimal situation to achieve power efficiency.

A possible solution to this inefficiency is to increase the noise shaping order of the VCO-ADC by employing what has been designated as "True VCO-ADC" architecture \cite{rombouts_true}. This architecture was first described in \cite{cardes-patente-vco} and later on, demonstrated in \cite{rombouts_true, rombouts-third-order,cardes-second-order}. Fig.~\ref{fig:comparison_prior_art} (b) depicts an example of a True VCO-ADC, that mimics the structure of a second-order CTSDM. In Fig.~\ref{fig:comparison_prior_art} (b), continuous time integrators are implemented with VCOs followed by digital counters that operate asynchronously \cite{vco_integrators}. The digital counters must count in two directions, the VCO increases the count at every cycle, while the feedback signal decreases the count by a variable amount corresponding to the output samples. Then, the loop filter is composed of a cascade of Digitally Controlled Oscillators (DCO) implemented as the combination of an asynchronous DAC driving a VCO (see Fig.~\ref{fig:comparison_prior_art} (b)). One of the advantages of this architecture compared to a conventional CTSDM is that DACs are in the feed-forward path of the loop filter and not in the feedback, relaxing the linearity requirements thanks to the loop gain. On the other hand, the input integrator sees the first VCO oscillation in open loop, being affected by VCO non-linearity but allowing the current reuse of the input preamplifier, while keeping high input impedance \cite{whyandhow,carlos_sensors,carlosmls,quintero_sscl}. In summary and as a main advantage, True VCO-ADCs allow to lower the effective frequency $f_e$ of the input VCO without compromising the Signal to Quantization Noise Ratio (SQNR) and thus improving power consumption. 

True VCO-ADCs are typically implemented with Ring Oscillators (RO), providing simultaneously with Pulse Frequency Modulation (PFM) encoding of the input signal and digital counting functions \cite{colorines1}. The counting function provided by ring oscillators results in a digital representation with thermometer coded values \cite{rombouts-third-order}. Thermometer coding apparently simplifies the interconnection logic in the loop filter of the sigma-delta modulator because add and subtract operations can be realized with an array of phase comparators \cite{rombouts-third-order}. However, thermometer coding represents a serious limitation of the dynamic range of state variables, which are constrained to be represented with the same number of levels than the number of VCO inverter stages. Solving this problem requires either limiting the input voltage swing \cite{rombouts-third-order}, which does not fully exploit the VCO gain or to increase exponentially the number of VCO taps, which results in a prohibitive area and ruins the efficiency of thermometer encoding. 

High performance digital microphones \cite{carlos_sensors,quintero_sscl}, usually require a large Dynamic Range (DR). Values in excess of 100 dB \cite{pavan_audio} measured with A-weighting (dB-A) are common in this application. A good design practice is to set the quantization noise a few dB below flicker and thermal noises. Clock frequencies for this type of audio ADCs are constrained by industry standards and lie below 4-5 MHz, which limits the oversampling ratio. These requirements translate, in the case of a first-order VCO-ADC, in a fairly large number of quantizer bits. As an example, in the first-order, open-loop ADC described in \cite{carlos_sensors}, an output word-length of 11 bits is used to reach 110 dB-A of DR. The large output word-length is implemented by a high frequency, multi-phase ring oscillator coupled to a coarse-fine counter architecture \cite{carlos_sensors,mercier_audio,steyaert_coarse-fine, vesterbacka}. As mentioned before, the quantizer word-length can be relaxed by using higher-order architectures \cite{vco-adc-perrott, rombouts-third-order, cardes-second-order, nan-sun-purely-vco-adc}. The True-VCO ADC topology \cite{rombouts-third-order, cardes-second-order} can be very attractive in this application if dynamic range can be extended, because the lack of a feedback DAC simplifies the direct coupling of the capacitive MEMS sensor to the VCO.

In this paper we propose a new kind of True-VCO-ADC modulators that use binary encoding and modulo arithmetic throughout all intermediate stages (See Fig.~\ref{fig:comparison_prior_art} (c)). This new approach enables a high dynamic range without requiring a large area. In the proposed architecture, the True-VCO-ADC topology is modified to be compatible with integer modulo arithmetic, rearranging the loop filter elements. Instead of using the RO phases as a thermometer encoded magnitude, the outputs of a ring oscillator are directly transformed into a Gray coded word \cite{gray_medina}. As a further advantage, inter-stage IDACs can be implemented using a space-saving R-2R architecture, given the relaxed linearity requirements. To prove experimentally the binary coded sigma-delta modulator, a 130nm - CMOS chip has been fabricated and measured. The chip implements a 103dB DR audio ADC which can directly interface with a capacitive MEMS microphone, operates at the standard sampling rate of 3.072MHz and occupies only 0.095 $\text{mm}^2$. The paper is organized as follows: Section II proposes a new nested True-VCO-ADC architecture and analyzes its binary coded implementation. Section III describes the system level design of the implemented ADC. Section IV describes the transistor level design of the chip. Section V discusses the experimental results and Section VI concludes the paper. 

\section{True VCO-ADC architectures with binary-weighted intermediate values}

Fig.~\ref{fig:comparison_prior_art}(a) and (b) describe a purely VCO-based ADC implementation, where all the integrator stages are composed by oscillators and digital logic. Fig.~\ref{fig:comparison_prior_art}(a) depicts a second-order modulator implemented with a VCO, a phase-frequency detector (PFD) and a Switched Ring Oscillator (SRO) described in \cite{nan-sun-purely-vco-adc}. In this VCO-ADC, an analog feedback signal is generated by a DAC around the VCO integrators. As an advantage, Fig.~\ref{fig:comparison_prior_art}(a) compensates the non-linearity of the VCOs because of the DAC feedback but then, the ADC is limited by the linearity of the same DAC. In addition, the input summation point represents a low impedance node which requires an additional input buffer if a high impedance source is to be used. Fig.~\ref{fig:comparison_prior_art}(b), depicts also a second-order modulator implemented with the True VCO-ADC architecture, where no explicit analog feedback DACs are required. Instead, feedback DACs are replaced by purely digital circuits, such as up-down counters, resulting in a mostly-digital architecture \cite{cardes-patente-vco}. 

For high order modulators, we can establish a trade off between quantizer resolution and modulator order. For instance, to reduce the output word-length of the VCO-ADC in \cite{carlos_sensors} from 11 bits to a mere 6 bits, it suffices to replace the first-order modulator by a second-order VCO-ADC. In these mostly-digital VCO-based ADCs, state variables are represented with a set of signals encoded with Pulse Width Modulation (PWM) and arranged as a as thermometer encoded magnitude \cite{rombouts-third-order}. Therefore bus widths and silicon area grow exponentially with the number of quantizer bits. Consequently, the area of the flip-flops and logic gates associated to these state variables will also increase exponentially. Assuming 6 bits are needed for second-order noise-shaping and with the interstage coupling circuits of \cite{rombouts-third-order}, the required number of flip flops and XOR/XNOR gates would be 320 and 128, respectively. Even with a third-order noise shaping modulator, the required number of flip-flops would be over 100, plus the extra VCO. Not only the area of the digital block increases exponentially with the number of bits, but also the VCO and subsequent Digital Controlled Oscillators (DCO) require a similar number of phases than the discrete levels of the state variables, which limits the design degrees of freedom. To cope with this limitation, the modulator in \cite{rombouts-third-order}, has a reduced input voltage range. The small input range translates into restricting the frequency variation of the first VCO to be properly represented with a reduced number of state variable levels. In the case of the modulator of \cite{cardes-second-order}, the sampling frequency is always larger than the VCO frequency, which allows the modulator to operate with single bit PWM coded state variables but at the expense of a clock frequency higher than the audio standards. 

\begin{figure}[t]
	\centering
	\includegraphics[width=\columnwidth,keepaspectratio]{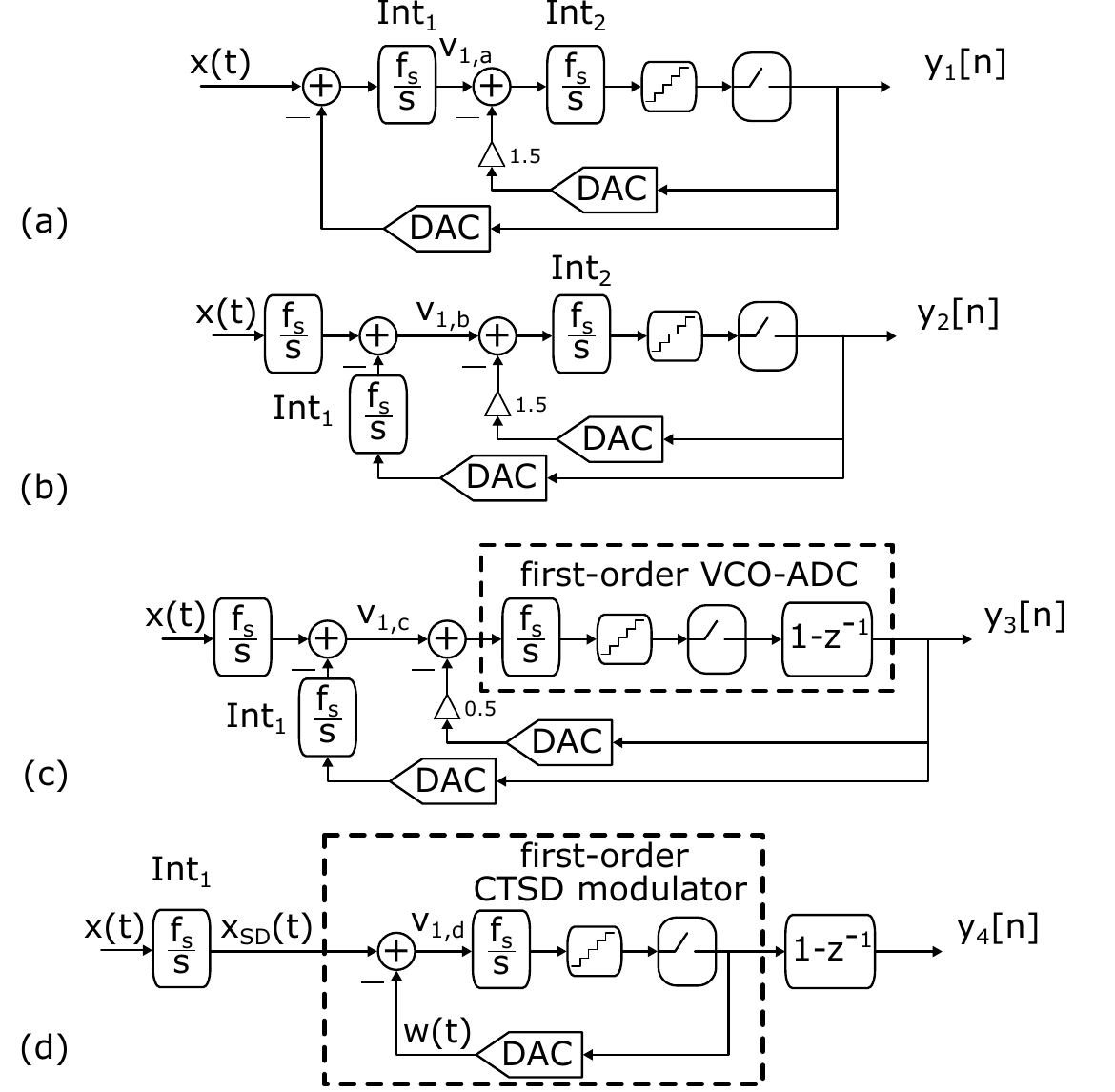}
	\caption{Different ways of building a second-order CTSD modulator, (a) typical CTSDM, (b) splitting input integrator, (c) substituting inner loop with a first-order VCO-ADC, (d) nested first-order CTSDM.}
	\label{fig:evolution_DS_to_Proposed}
\end{figure}

As a contrast, Fig.~\ref{fig:comparison_prior_art}(c) shows our proposed architecture that enables a binary weighted implementation of a True VCO-based ADC.  In our proposed architecture, the number of digital elements, and consequently the area, increase linearly with the number of bits instead of exponentially. To derive the architecture of Fig.~\ref{fig:comparison_prior_art}(c), we will start with a conventional second-order CTSDM that will be modified applying the linear transformations described in Fig.~\ref{fig:evolution_DS_to_Proposed}. The block diagram of this second-order CTSD modulator is shown in Fig.~\ref{fig:evolution_DS_to_Proposed}(a) and has a $\text{NTF}=(1-z^{-1})^{2}$ and $\text{STF}=\text{sinc(f)}^2$. We take the first integrator ($Int_1$) out of the loop and place one integrator on the input and another one on the outer feedback path as Fig.~\ref{fig:evolution_DS_to_Proposed}(b) shows. Then, we can replace the first-order CTSDM built around the second integrator, $Int_2$, by a VCO-based ADC, which is composed by a VCO acting as phase integrator, a phase quantizer, a sampler and a first difference block (See Fig.~\ref{fig:evolution_DS_to_Proposed}(c)). This replacement has been shown to be mathematically identical in \cite{CASM_CTSDM_PFM}. In order to keep the same NTF as in Fig.~\ref{fig:evolution_DS_to_Proposed}(a), an extra feedback loop with gain 0.5 around the first-order VCO-ADC is needed, as shown in Fig.~\ref{fig:evolution_DS_to_Proposed}(c). We observe that after sampling is done, the feedback branch connected to the outer feedback loop includes a discrete-time first-difference block that is cancelled immediately afterwards by a continuous-time integrator.  Thus, a simplification can be made, collapsing the first difference and the integrator in the feedback path. Therefore, the feedback path can be taken right after the sampler, pushing the first difference outside of the loop. The resulting NTF from Fig.~\ref{fig:evolution_DS_to_Proposed}(d) corresponds to the first difference block at the end multiplied by the NTF of a fist-order CTSD Modulator which also is $1-z^{-1}$. Therefore, Fig.~\ref{fig:evolution_DS_to_Proposed}(d) has the same NTF than Fig.~\ref{fig:evolution_DS_to_Proposed}(a). To achieve this similarity, the extra path around the VCO-ADC in Fig.~\ref{fig:evolution_DS_to_Proposed}(c) has to be removed. This also results in the same STF.

\begin{figure}[t]
	\centering
	\includegraphics[width=\columnwidth,keepaspectratio]{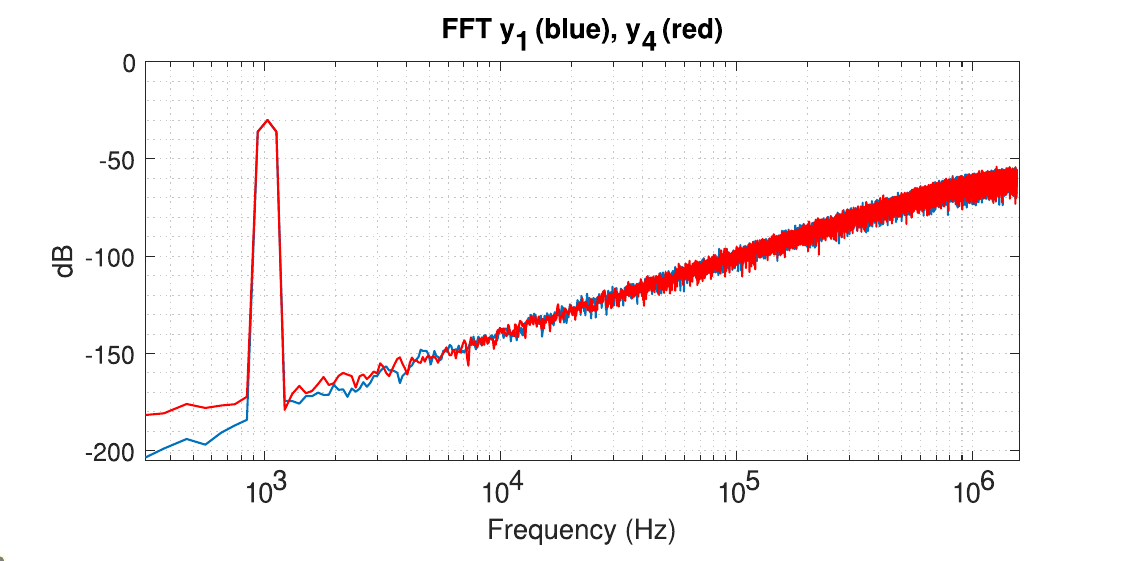}
	\caption{Average of 32 FFT from system level simulations of Fig.~\ref{fig:evolution_DS_to_Proposed}(a) and Fig.~\ref{fig:evolution_DS_to_Proposed}(b).}
	\label{fig:sim_simple_complete}
\end{figure}

\begin{figure}[t]
	\centering
	\includegraphics[width=\columnwidth,keepaspectratio]{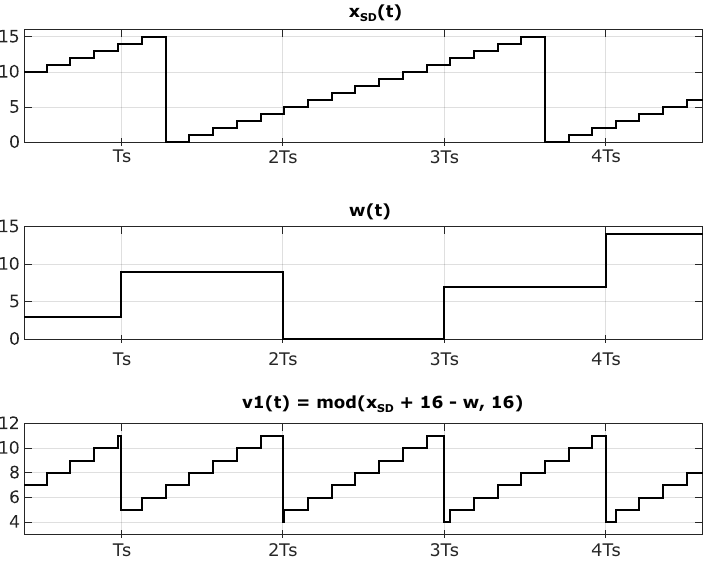}
	\caption{Operation of the subtraction element SB with modular arithmetic.}
	\label{fig:wrapping_mechanism}
\end{figure}

The final equivalent CTSD modulator results simpler than the original CTSDM, as can be seen in Fig.~\ref{fig:evolution_DS_to_Proposed}(d). This final CTSDM is composed of an integrator, followed by a first-order CTSDM and a first difference. This simple model could not be practically implemented using voltage or current based-integrators as node $x_{SD}(t)$ in Fig.~\ref{fig:evolution_DS_to_Proposed}(d) would be an unbounded magnitude. However, if both integrators are implemented with VCOs \cite{cardes-second-order, rombouts-third-order}, the modulator can be designed such that the feedback signal $w(t)$ compensates variable $x_{SD}(t)$ yielding an always bounded difference. If we assume that $Int_1$ is implemented as a VCO and a finite word-length counter with M levels, $x_{SD}(t)$ will be an integer number wrapping every M counts. If $Int_2$ is also implemented as a VCO followed by a finite word-length counter, $w[n]$ will be also bounded and the difference will correctly represent state variable $v_{1,d}(t)$ under some conditions. We first need that the subtraction operation is implemented in modulo arithmetic, as in a CIC filter \cite{Hogenauer} and also that $x_{SD}(t)$ does not wrap more than once every sampling period, or there will be a saturation in the state variable. State variables $v_1$ in Fig.~\ref{fig:evolution_DS_to_Proposed}, $v_{1,a}$, $v_{1,b}$ and $v_{1,c}$ are all the same, however $v_{1,d}$ is different as the contribution from the feedback path does not pass through an integrator. Fig. \ref{fig:sim_simple_complete} shows a comparison of a behavioural simulation, with different inital conditions, of Figs.~\ref{fig:evolution_DS_to_Proposed} (a) and (d), proving the equivalence. Using the behavioral model of Fig. \ref{fig:evolution_DS_to_Proposed}, we have represented in Fig. \ref{fig:wrapping_mechanism} an example of the operation of the substractor after $Int_1$, with M=16. The value of $v_{1,d}(t)$ is defined by the following equation:

\begin{eqnarray}
	v_{1,d}(t)=mod_{M}(x_{SD}(t)+M-w(nT_s))  \label{eq:e_calculation}
\end{eqnarray}

\begin{figure}[t]
	\centering
\includegraphics[width=\columnwidth,keepaspectratio]{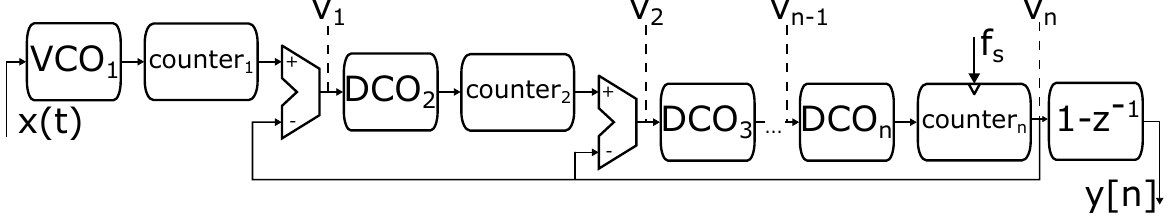}
	\caption{Extension to higher order modulators.}
	\label{fig:higher_order_example}
\end{figure}

This same principle can be applied to any modulator implemented as a cascade of integrators with multiple feedback. Fig.~\ref{fig:higher_order_example} shows this kind of generic True-VCO ADC using Digitally Controlled Oscillators (DCO), binary counters as integrators and  digital substractors using modulo arithmetic to implement the feedback connections. Signals $v_1, v_2...v_N$ are asynchronous digital numbers expressing the state variables of this kind of loop filter in modulo arithmetic. 

\section{Architecture analysis of the proposed $2^{nd}$ order VCO-ADC} \label{sec:system-design}
In this section we describe at the system level the second-order VCO-ADC architecture used in our experimental chip. We provide a detailed explanation of the new blocks and calculate a linearized model of the system.

\subsection{Block Diagram of Proposed Architecture}
\begin{figure*}[!t]
	\centering
\includegraphics[width=2\columnwidth]{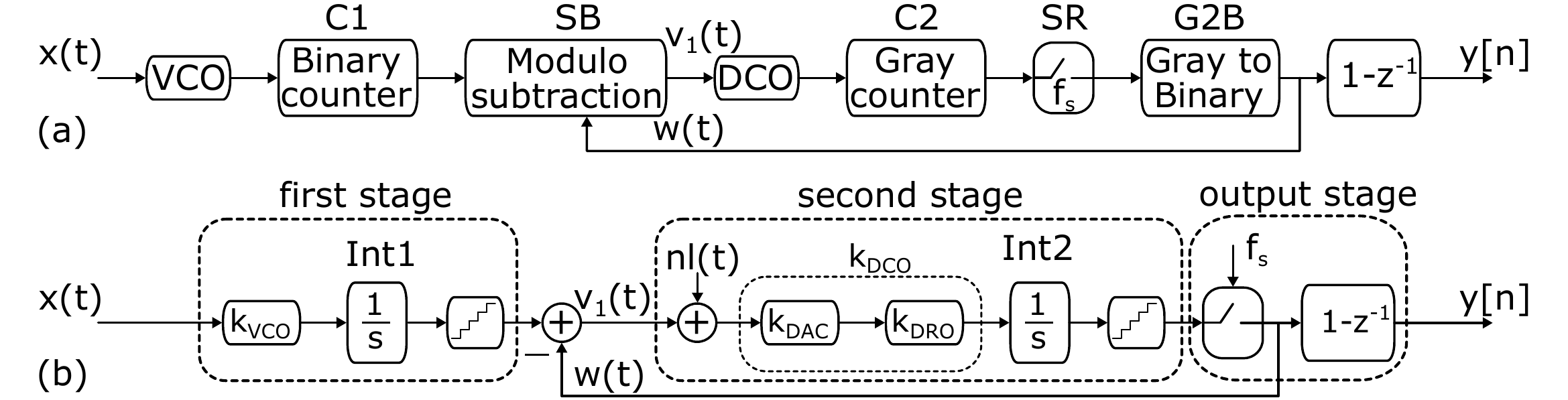}
	\caption{(a) Proposed system and (b) its simplified linear model.}
	\label{fig:system}
\end{figure*}

Building upon the models of the previous section, Fig.~\ref{fig:system} presents a detailed conceptual block diagram of our implementation. The input signal $x(t)$ modulates the input VCO implemented as a multi-phase RO. The VCO output pulses are accumulated in a synchronous binary counter $C1$. This counter can only have 16 values. Thus the input signal to the loop will indefinitely keep wrapping. To cope with the wrapping mechanism, a digital modulo subtractor $SB$ is used as the feedback node, as described in Section II.
As long as the loop is able to follow the input (like in a locked digital PLL), signal $v_1(t)$ will be equivalent to that of using ideal boundless counter in $C1$. This signal $v_1(t)$, modulates a DCO, implemented with a Digital to Analog Converter (DAC) driving a differential RO. The pulses of this DCO need to be accumulated in a second counter $C2$, whose output is periodically sampled by a digital register $SR$ at the modulator sampling rate. This sampling operation poses some difficulties as the DCO and the sampling clock are asynchronous. Therefore, metastability errors can affect the sampled values. If we use a binary counter for $C2$, metastability errors due to sampling are not limited to 1 LSB because in a binary code, all bits can switch at once. This problem is present regardless if counter $C2$ is built with an asynchronous or a synchronous architecture. Consequently, unless a mechanism is used to limit metastability errors, the sampled sequence will present glitches and the SQNR will be diminished. To overcome this problem, the approach in our proposed architecture is to use as $C2$ a Gray coded counter. In \cite{gray_medina} it is shown that is possible to use the phases of a differential RO in order to directly generate a Gray coded sequence. Gray coded counters have been used \cite{quintero_sscl,vesterbacka} to limit the metastability sampling error to 1 LSB due to the properties of Gray code. The sampled signal $w[n]$ is the binary feedback to the modulo subtraction block. Therefore, Gray code coming from $C2$ must be decoded into a binary word by block $G2B$. Finally, this binary output $w[n]$ is also connected to a first difference block, yielding the decoded output signal, $y[n]$ which presents second-order noise-shaping.

\subsection{Gray Encoded Ring Oscillator}
\begin{figure}[t]
	\centering
	\includegraphics[width=\columnwidth,keepaspectratio]{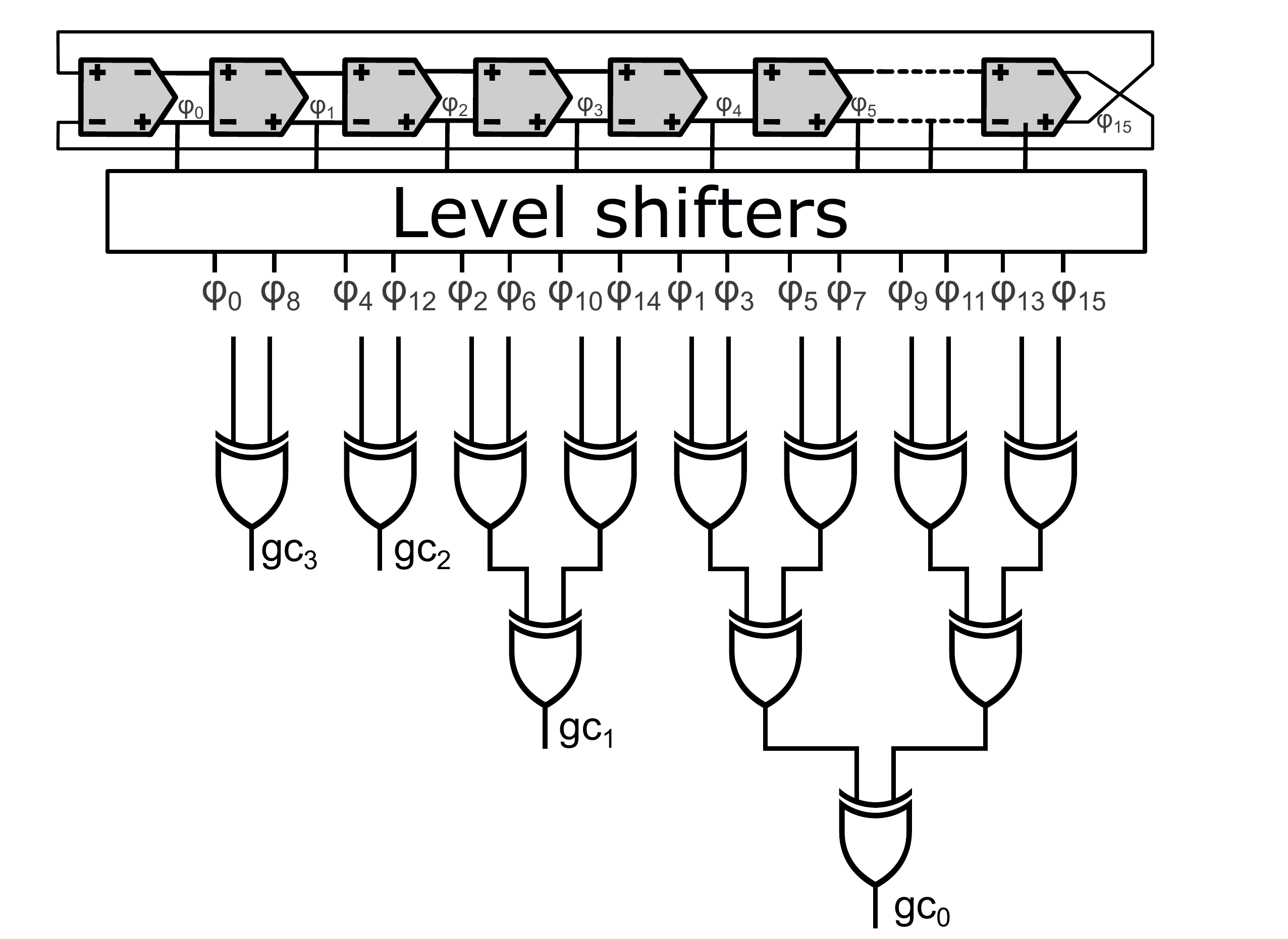}
	\caption{Gray counter directly connected to the phases of a differential ring oscillator.}
	\label{fig:XOR_grayEncoding}
\end{figure}

Based on \cite{gray_medina}, we propose an improved Gray encoding architecture that allows direct connection to the DCO. Fig.~\ref{fig:XOR_grayEncoding} shows, for simplicity, an implementation example with 4 bits which includes a 16 tap differential RO, a group of 16 level shifters to bring the RO signals to a valid logic level and a digital encoder. The digital encoder is composed only by XOR gates. In a Gray code, the switching frequency of the bits doubles every bit, as we move from the LSB to the MSB. In contrast to binary encoding, in Gray encoding, the last 2 MSB have the same switching frequency. Thus, as shown in Fig.~\ref{fig:XOR_grayEncoding}, in order to generate the 2 MSB, $gc_{3}$ and $gc_{2}$, only 1 XOR gate for each bit is needed. For this example, where the RO has 16 delay stages, the number of XOR gates needed is 12. This architecture requires less XOR gates than the number of RO phases and no Flip Flops (FF), as an advantage compared to \cite{gray_medina}. Furthermore, because output $gc$ is in base 2, the number of sampling FF required is the same as the number of bits (4 in this case). The relationship between the digital area and the number of bits is therefore linear. The connection order from the RO to the XORs is critical, as it is what yields the correct Gray code sequence. Let the RO oscillate at frequency $f$, then the resulting Gray encoding sequence is a set of digital signals with a 50\% duty cycle. In this example the LSB, $gc_{0}$ bears the fastest rest frequency at $16f$, as it switches once every two consecutive states of the ring oscillator. From this point, bits $gc_{1}$ to $gc_{3}$ switch at slower frequencies. Therefore, $gc_{0}$ defines the required speed for the digital logic. The digital encoder can be described by the following equations:

\begin{eqnarray}
	gc_{0}= \varphi_{1}  \oplus  \varphi_{3}  \oplus  \varphi_{5}  \oplus  \varphi_{7}  \oplus  \varphi_{9}  \oplus  \varphi_{11}  \oplus  \varphi_{13}  \oplus  \varphi_{15}  \nonumber \\ 
    gc_{1}= \varphi_{2}  \oplus  \varphi_{6}  \oplus   \varphi_{10}  \oplus  \varphi_{14}\nonumber \\ 
    gc_{2}= \varphi_{4}  \oplus   \varphi_{12} \\ \nonumber
    gc_{3}= \varphi_{8}  \oplus  \varphi_{0} 
    \label{eq:phases_DCO_for_gc_4bits}
\end{eqnarray}
In a general case with N bits, we can propose the following algorithm to define the equations of the digital encoder: 
\begin{eqnarray}
	gc_{n}= P(\varphi_{k}) \, , \,  k = 2^{n} \cdot (2j-1) \, , \, j=1...{\frac{N+1}{2^{n+1}}} \nonumber \\ 
    gc_{N-2}= \varphi_{N/4}  \oplus  \varphi_{N/2+N/4}
    \label{eq:general_phases_DCO_for_gci}
\end{eqnarray}
\begin{eqnarray}
    gc_{N-1}= \varphi_{N/2}  \oplus  \varphi_{0} 
    \label{eq:lastBit_DCO_for_gci}
\end{eqnarray}
where $\oplus$ is the XOR operator and $P(\cdot)$ represents the Parity function (joint XOR operation of a series of single-bit logic variables). All bits are represented by \eqref{eq:general_phases_DCO_for_gci} except the $gc_{N-1}$, as its frequency is the same as $gc_{N-2}$. Bit $gc_{N-1}$ is described by \eqref{eq:lastBit_DCO_for_gci}. This new architecture makes use of both rising and falling edges of the RO, which is an advantage with respect to previous work \cite{gray_medina}, as it allows to have a lower RO rest frequency without degrading SQNR. It also allows the reduction of current consumption while keeping the same flicker noise \cite{abidi-noise_CMOS_RO, borgmans-noise-RO}. In comparison with \cite{gray_medina} there are no FF and therefore there is no need for a start-up circuit. If \eqref{eq:general_phases_DCO_for_gci} and \eqref{eq:lastBit_DCO_for_gci} are used for the connection of the phases, the output Gray code will be correct, no matter how the initial transient of the oscillator. However, in a differential oscillator there is one delay stage whose output only delays with respect to the previous one, without inverting it. As this is critical, when connecting the phases this should not be disregarded. For simplicity, this phase is considered to be $\varphi_0$, even though other indexes are also suitable. By reordering the RO phases, a downwards counting sequence is also possible.

\subsection{Linear model of the proposed architecture}
Fig.~\ref{fig:system}(b) shows a simplified linear model of our implementation that follows the architecture described in Fig.~\ref{fig:evolution_DS_to_Proposed}(d). The VCO and DCO in Fig.~\ref{fig:system}(a) are represented by two integrators with gains $k_{DCO}$ and $k_{VCO}$ respectively. The DCO block is implemented as a DAC with gain $k_{DAC}$ followed by a RO with gain $k_{DRO}$, therefore $k_{DCO}=k_{DAC} \cdot k_{DRO}$. After every integrator we have placed a quantizer that represents the discrete values generated by counters $C1$ and $C2$ in Fig.~\ref{fig:system}(a). Both quantizers are considered to have the same quantization step. The contents of $C2$ are sampled at the sampling frequency $f_s$ and used as feedback signal. Finally the output sequence $y[n]$ is generated with a first difference block. 

 There are several conditions for the model of Fig.~\ref{fig:system}(b) to accurately represent the proposed architecture of Fig.~\ref{fig:system}(a). First of all, the gain $k_{DCO}$ and $k_{VCO}$ should be approximately of the same value to allow locking of first and second stages. They are also considered to be the effective values (referenced to $f_e$). A mismatch between both oscillator rest frequencies reflects as a decrease of the dynamic range. In Fig.~\ref{fig:system}(a), the finite number of bits and the wrapping mechanism of $C1$ and $C2$ may result in an overflow after the subtraction in SB. As a difference, the ideal system of Fig.~\ref{fig:system}(b) does not reflect this overflow, given the unbounded operation of the integrators. As a consequence of the overflow, there is a point where $C2$ might not \textit{lock} to $C1$ if the input signal drives the input VCO into a frequency out of full scale. 

Using the linear model of Fig.~\ref{fig:system} (b) we can estimate the NTF of the modulator. The feedback gain is set to 1 and the digital implementation allows it to be independent of non-idealities, and it is therefore constant. A general expression of the NTF, that takes into account any arbitrary value of the integrator gains with respect to $f_s$, can be written as follows:

\begin{eqnarray}
	\text{NTF}(z)=\frac{1-z^{-1}}{1-(1-k_{DCO}/f_s)\cdot z^{-1}} \cdot (1-z^{-1}).  \label{eq:NTF_general}
\end{eqnarray}

As an advantage of our architecture compared to a standard CTSD modulator, we may point out the insensitivity to Process, Voltage and Temperature (PVT) variations. First, the  gain of the input VCO, $k_{vco}$, only affects the signal gain and not the positions of NTF poles and zeros, considering that the feedback is a digital subtraction. The first difference placed at the output has a digital implementation, therefore it will not be affected by PVT. The NTF presents second-order noise shaping with, at most, one pole. The pole of the resulting NTF only depends on the $k_{DCO}$. In contrast, any deviation in the gain of the second integrator of a conventional CTSD, may produce complex-conjugated poles. Thus, this system presents second-order noise shaping but the dynamics of the system are the same as a first-order CTSD modulator. For instance, the amount of Excess Loop Delay (ELD) that the system can tolerate before becoming unstable is that of a first-order modulator. Most of these properties are common to other VCO-based higher-order architectures \cite{cardes-second-order, rombouts-third-order, nan-sun-purely-vco-adc, vco-adc-perrott}. The STF can also be derived form Fig.~\ref{fig:system}(b) as follows:

\begin{eqnarray}
	\text{STF}= \frac{k_{vco}}{fs} \cdot sinc(f)^2
 \label{eq:STF_general}
\end{eqnarray}

Any gain mismatch in $k_{DCO}$ will not significantly affect the in-band signal gain of the STF. The STF of the whole system results from the combination of the input integrator, first difference, and the STF of the loop in Fig.~\ref{fig:system}(b). We can assume that there is little-to-none excess-loop-delay (ELD) in the feedback loop due to its digital implementation. Therefore, the resulting STF will be completely flat within the band of interest, followed by a second-order roll-off attenuation. As in any open loop VCO-based ADC, the gain of the STF mainly depends on the $k_{vco}$ and $f_s$, as can be seen in \ref{eq:STF_general}.

A circuit impairment reflected in Fig.~\ref{fig:system}(b) are the DAC nonlinear components, that we have represented as the additive signal $nl(t)$. We recall that the modulo subtraction block $SB$ of Fig.~\ref{fig:system}(a) operates in binary and therefore the bits of $e(t)$ do not rotate like in the thermometer encoded implementation of \cite{rombouts-third-order}. Therefore, there is no intrinsic data weight averaging (DWA) \cite{rombouts-third-order} or circuit-level-averaging \cite{nan-sun-purely-vco-adc}. Nonetheless, we can observe that in our proposed system and in \cite{rombouts-third-order}, the possible non linearity errors $nl(t)$ are introduced at the input of the second integrator $Int2$, as a difference to \cite{nan-sun-purely-vco-adc} whose DAC is directly subtracted form the input. Even though our system does not have DWA, observing Fig.~\ref{fig:system} (b) we see that $nl(t)$ is first-order shaped. Therefore, nonlinearity components inserted at $Int2$ will be attenuated by approximately 55 dB at the end of the 20KHz BW.  

\begin{figure*}[!t]
	\centering
	\includegraphics[width=2\columnwidth,keepaspectratio]{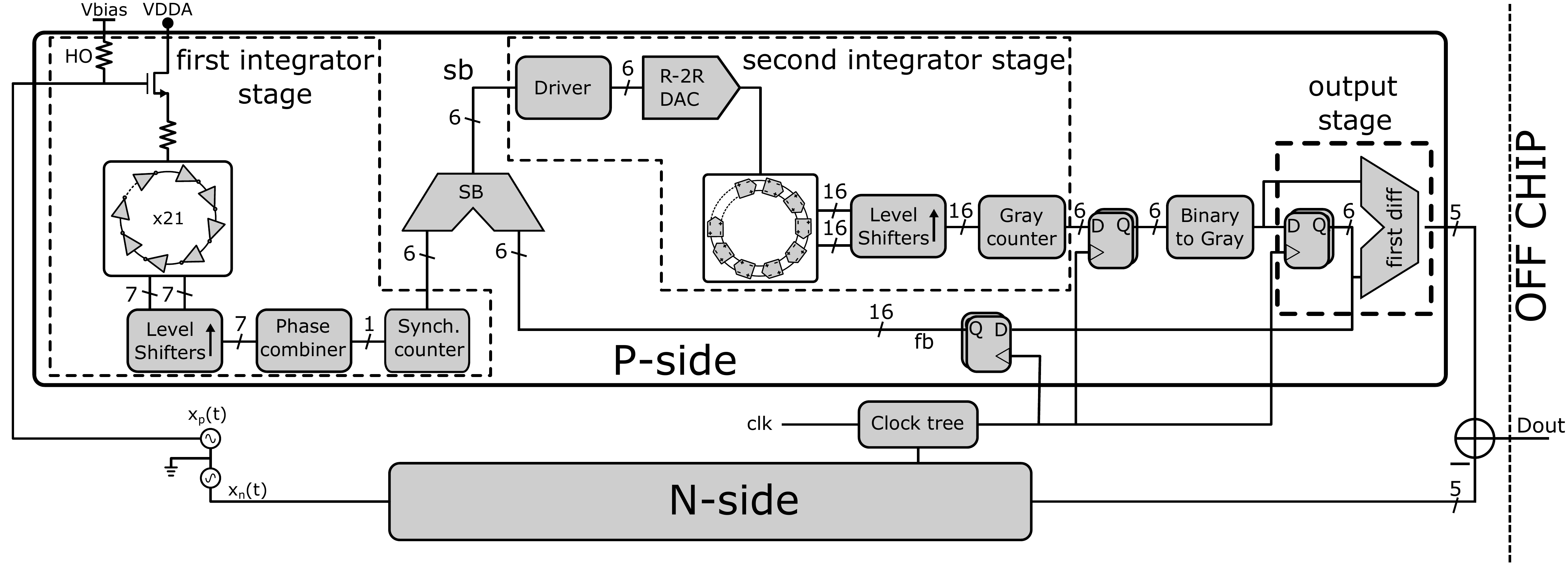}
	\caption{Circuit level implementation of the proposed second-order True VCO-ADC.}
	\label{fig:complete_system}
\end{figure*}

\begin{figure}[t]
	\centering
	\includegraphics[width=\columnwidth,keepaspectratio]{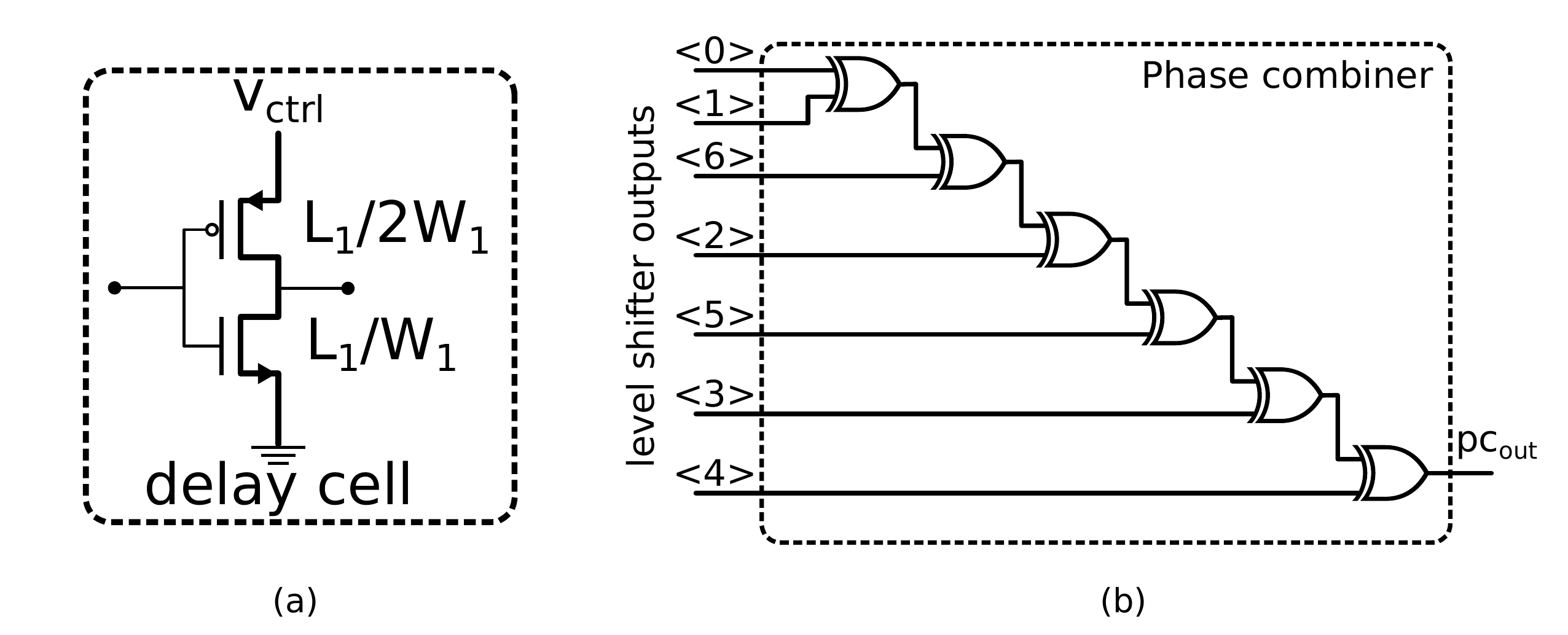}
	\caption{(a) Delay stage of the first stage. (b) circuit of the phase combiner.}
	\label{fig:circuit_first_stage}
\end{figure}

\section{Circuit level design} \label{sec:circuit-design}
In this section we will describe the circuit implementation, shown in Fig.~\ref{fig:complete_system}, of our proposed architecture. This circuit closely follows the block diagram depicted in Fig.~\ref{fig:system}(a). Since we use a pseudo-differential architecture, there is a P-side and N-side channels receiving differential input signals, $x_p(t)$, $x_n(t)$. Fig.~\ref{fig:complete_system} depicts the details of the P-channel only as both channels are identical. Each channel is composed of a first integrator stage, cascaded with a second integrator stage. Finally, an output stage computes the first difference of each channel and a subtractor provides the ADC output, $D_{out}$.

\subsection{First Integrator Stage}
The input signal $x_p(t)$ is connected to the first integrator stage (see dashed line in Fig.~\ref{fig:complete_system}). The first stage is composed of a Source Follower (SF) whose load is a resistor in series with a VCO. The bias voltage at the gate of the SF is provided through a high ohmic resistor (HO) and a external voltage source to provide high input impedance. The input voltage $x(t)$ is meant to be generated by a constant-charge biased capacitive MEMS microphone \cite{luca-sant_mems_microphone}. Therefore the input signal can be represented by a source of very high impedance that is capacitive coupled to the SF gate. The resistor in series with the VCO helps to linearize the VCO by providing negative feedback in the node $v_{ctrl}$ \cite{rombouts_electronic_letters}. A side effect of this linearization resistor is to lower the $k_{vco}$, but brings the benefit a higher Signal-to-Noise-Distortion-Ratio (SNDR). 

The VCO is implemented as a RO composed of single-ended CMOS inverters as delay stages and is shown in Fig.~\ref{fig:circuit_first_stage} (a). According to Section~\ref{sec:system-design}, the required effective frequency of the RO should be $f_e=42~MHz$ to achieve $>$100 dB of DR for a second-order system. The supply of the RO is modulated according to $v_{ctrl}$ and thus its output phases are also amplitude modulated (see Fig.~\ref{fig:circuit_first_stage}(a)). Consequently, level shifters (LS) \cite{lanuzza_levelshifters, quintero_sscl} are used in order to eliminate the amplitude modulation and convert the output phases to valid \textit{digital} levels. Counter C1 in Fig.~\ref{fig:complete_system} is a 6-bit, synchronous binary counter triggered by rising edge. To couple the RO to C1 a phase combiner circuit (PC), (see Fig.~\ref{fig:circuit_first_stage}(b)) is used to extract the effective frequency $f_e$ from the RO phases. The PC is composed of a chain of XOR gates. The 7 inputs to the PC are the LS outputs. The maximum frequency for the PC output $pc_{out}$ and the considered input range is not higher than $100$ MHz. The phases selected for the PC must be equidistant such that the frequency at the output of the phase combiner is effectively increased and so does the SQNR \cite{medina_analysis_vco_subset_feedback)}.  

To optimize power and area of this block, several design decisions must be made to dimension the RO and its associated circuitry. We have chosen a rest frequency $f_0=6$~MHz and 21 inverters for the RO. From the possible 21 output phases we will select only $M_1=7$ phases to generate the counter C1 input ($f_e=$7~Phases $\times$  6~MHz$=$42~MHz). These 7 RO phases are connected to the active level shifters, while the remaining unused 14 phases are connected to dummy level shifters to balance the load seen by all inverters. The reasoning for this decision is as follows. A 7 tap, smaller oscillator, could have been designed running at the same $f_e$ and avoiding unused outputs. However, that solution, apparently advantageous, has several drawbacks. In our design process we aimed to decrease flicker noise such that the RO noise is dominated by thermal noise while using a relatively low current. Then, the proposed solution with 21 phases has resulted in one of the most efficient solutions in power and area to achieve our target noise budget. Our design consumes 55~$\mu$A per each branch and is dominated by thermal noise. If the 21 phases were used, level shifters and subsequent digital circuitry would significantly increase power and area. Note that using all 21 phases would increase the effective $f_e$ in the VCO, which implies increasing the $f_e$ of the DCO in the second stage which further increases power consumption and digital area. This way power efficiency is guaranteed by limiting the ADC by thermal and flicker noises instead of quantization noise. 

To implement C1, we have used a 6-bit synchronous binary counter in which all the output bits, $bc$, switch with the same clock signal, $pc_{out}$. We have preferred a synchronous counter instead of a more power efficient asynchronous counter to reduce propagation delays. However, the delay between $bc$ and $pc_{out}$ is not very relevant, since this delay does not affect SQNR in any way and can be treated as a delay on the input signal $x(t)$. In Fig. \ref{fig:complete_system}, substractor SB is implemented as 6 full adders connected with a carry chain.

\begin{figure}[t]
	\centering
	\includegraphics[width=\columnwidth,keepaspectratio]{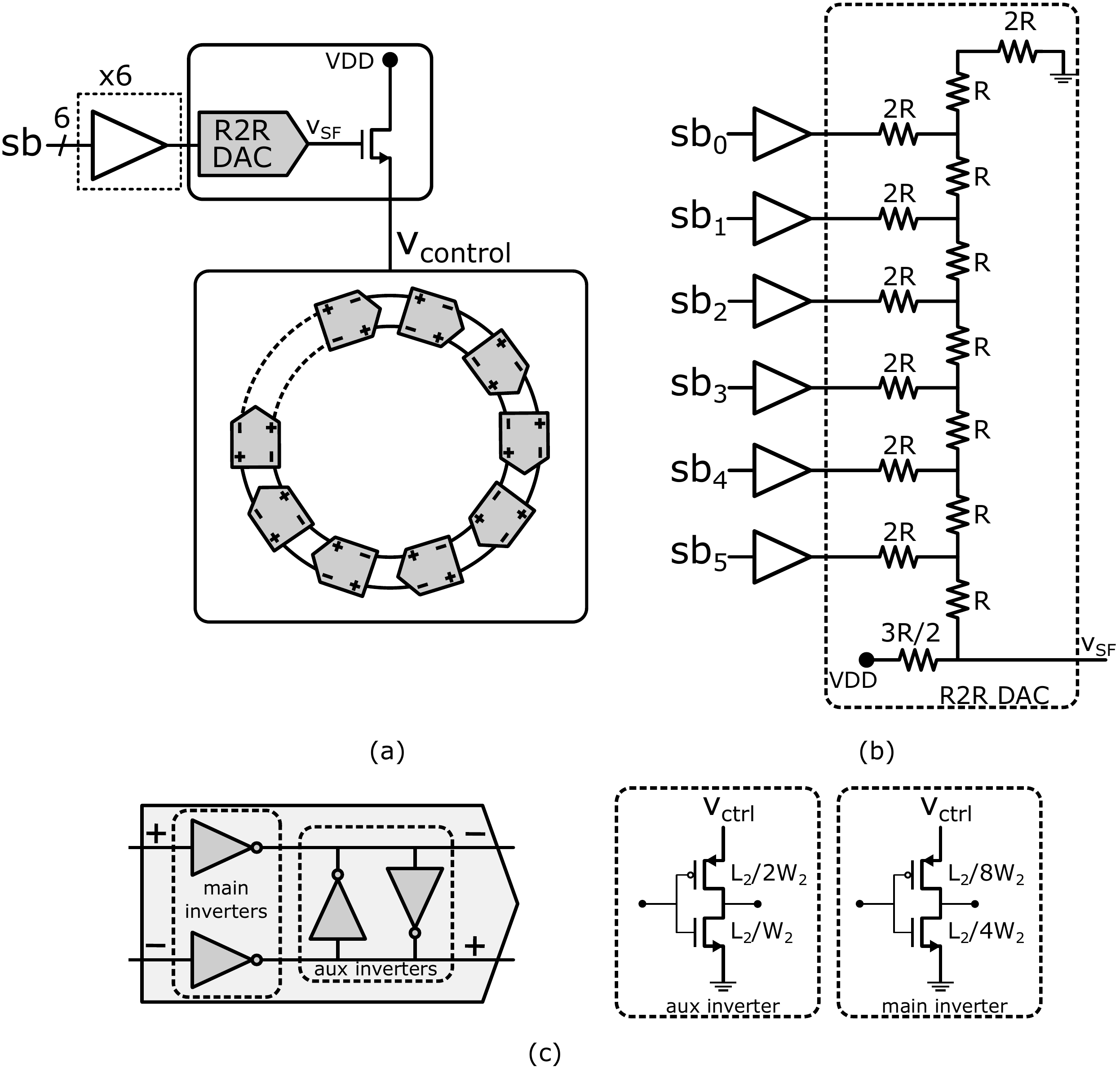}
	\caption{DCO circuit.}
	\label{fig:circuit_second_stage}
\end{figure}

\begin{figure}[t]
	\centering
	\includegraphics[width=\columnwidth,keepaspectratio]{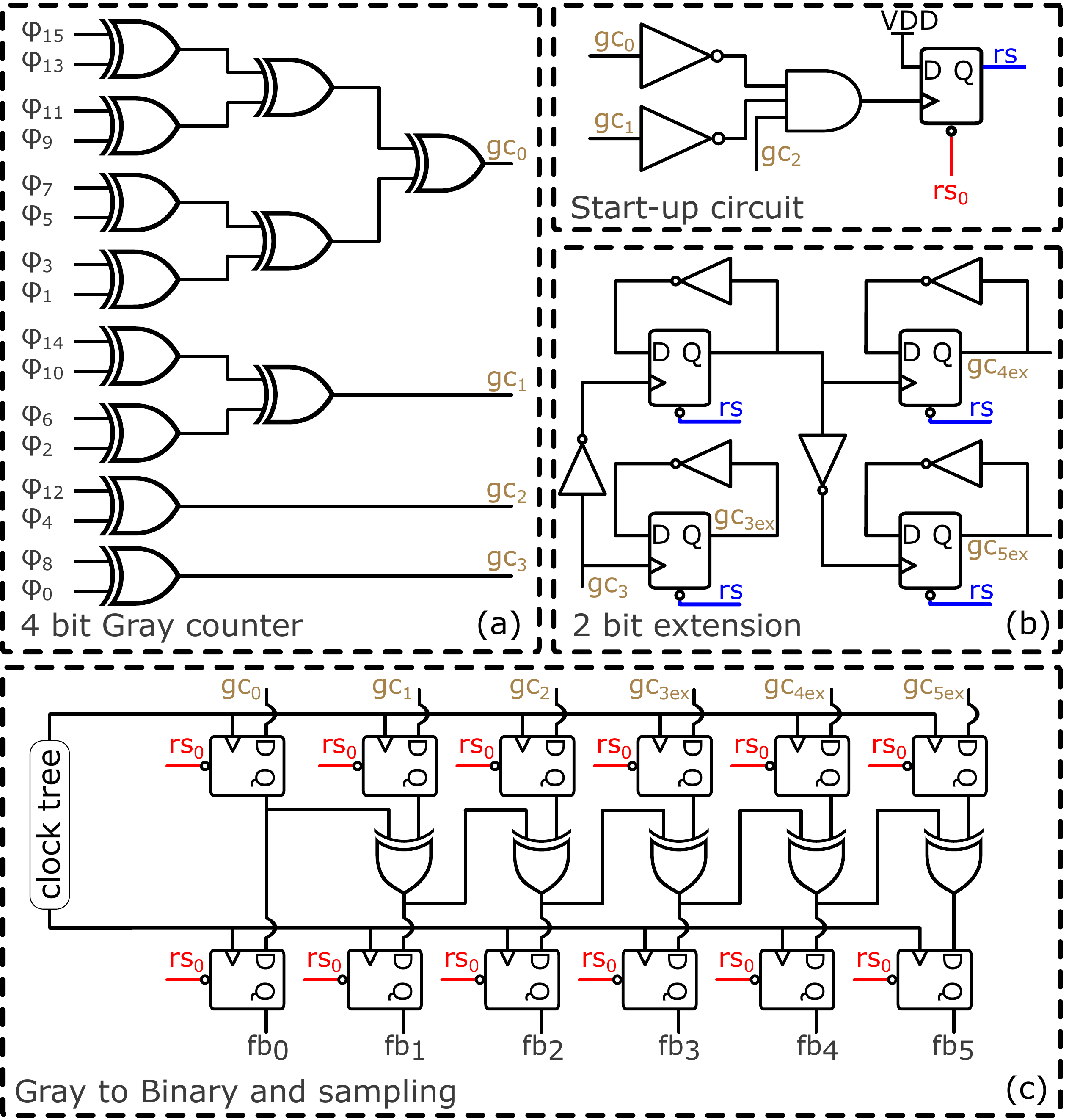}
	\caption{Circuit details of 6-bit counter, (a) 4-bit Gray counter directly connected to the RO phases, (b) Circuit to extend the Gray count to 6-bit, (c) Sampling and Gray to binary decoder.}
	\label{fig:DCOcounter_sampling}
\end{figure}

\subsection{Second Integrator Stage}
Fig.~\ref{fig:complete_system} shows in dashed line the second integrator stage. The implementation of the DCO is shown in Fig.~\ref{fig:circuit_second_stage}(a), which comprises the DAC drivers, a R-2R six bit DAC and a differential RO with 16 taps. The number of states of the RO must match with the number of Gray-encoded values at the output of the second integrator. Therefore, the RO needs a power-of-two number of taps. An area saving choice would have been a RO with single-inverter taps. However, single inverter ROs must have an odd number of phases. Unevenly distributed phases degrade the SQNR, as described in \cite{medina_analysis_vco_subset_feedback)}, and therefore we have preferred the differential RO. 

We have designed a differential ring-oscillator composed of 16 direct cross-coupled delay cells (see Fig.~\ref{fig:circuit_second_stage}(c)). Even though the noise performance is slightly worse in comparison to a feed-forward cross-coupled delay cell \cite{borgmans-noise-RO}, the increase in frequency provided by the latter is actually not desirable in our approach, given the low required rest frequency $f_0=1.2MHz$. The output of the differential RO is connected to level shifters to obtain digital levels for the subsequent digital logic. To drive the differential RO, a simple R-2R ladder DAC plus a SF voltage buffer is used (see Fig.~\ref{fig:circuit_second_stage}(b)). The SF isolates the impedance of the DAC and the differential RO. Because the DAC is in the middle of the Sigma-Delta loop, the thermal noise of the DAC resistors is attenuated. The selected value for the poly-resistors is $R=100$ k$\Omega$. Also, poly-resistor flicker noise is suppressed by the loop, and consequently, a moderate-size unitary resistor can be used. The total active area of the implemented R-2R ladder is of 485 $\mu$m$^2$. The SF of the DCO presents a similar area. The whole second integrator stage is powered by the digital supply, given the noise attenuation of the loop.

Due to the DCO design choice of 16 taps, the Gray counter described in Section~\ref{sec:system-design} has 16 different states only (4 bits) but must generate a 6 bit word at the input of the sampler. Therefore an scaling circuit is needed before the sampler. Fig.~\ref{fig:DCOcounter_sampling} shows the implemented circuit for the Gray counter and the scaler. Bits $gc_{0}, gc_{1}, gc_{2}$ coming from the DCO Gray counter in Fig.~\ref{fig:DCOcounter_sampling}, are directly connected to the sampler. The fourth bit, $gc_{3}$ is used to generate bits bits $gc_{3ex}, gc_{4ex}$ and $gc_{5ex}$, the remaining 3 bits to complete the 6 bit, Gray-encoded word that is actually sampled. The circuit to generate the extra bits is depicted in detail in Fig.~\ref{fig:DCOcounter_sampling}. Fig.~\ref{fig:DCOcounter_sampling}(b) shows the start-up circuit: after the start of the whole chip, the digital circuits are reset using $rs_{0}$ which then generates $rs$ that is used to generate the extended bits coherently.

\subsection{Output Stage}
The 6-bit Gray-encoded output bus is sampled in a register at the sampling clock of $f_s=3.072$~MHz. After sampling, the Gray-coded values are decoded into binary words with a chain of XOR decoder and resampled to compensate the propagation delay introduced by the Gray-to-Binary decoder tree. This 6-bit binary signal is fed back into block SB. To synchronize the feedback branch and the output stage without incurring in significant excess loop delay, the synchronization circuit and clock tree shown in Fig.~\ref{fig:complete_system} have been used. Finally, Fig. \ref{fig:complete_system} includes a 6-bit first-difference circuit for each branch. However, only 5 bits are taken at the output of the first difference block, because the 6-th bit is only allocated to avoid overflow in the inner loop (see Fig.~\ref{fig:wrapping_mechanism}). The 6-bit output sequence $D_{out}$ is generated by a subtractor that computes the differential component between both 5-bit branches.  

\section{Experimental Results}
\begin{figure}[t]
	\centering
	\includegraphics[width=\columnwidth,keepaspectratio]{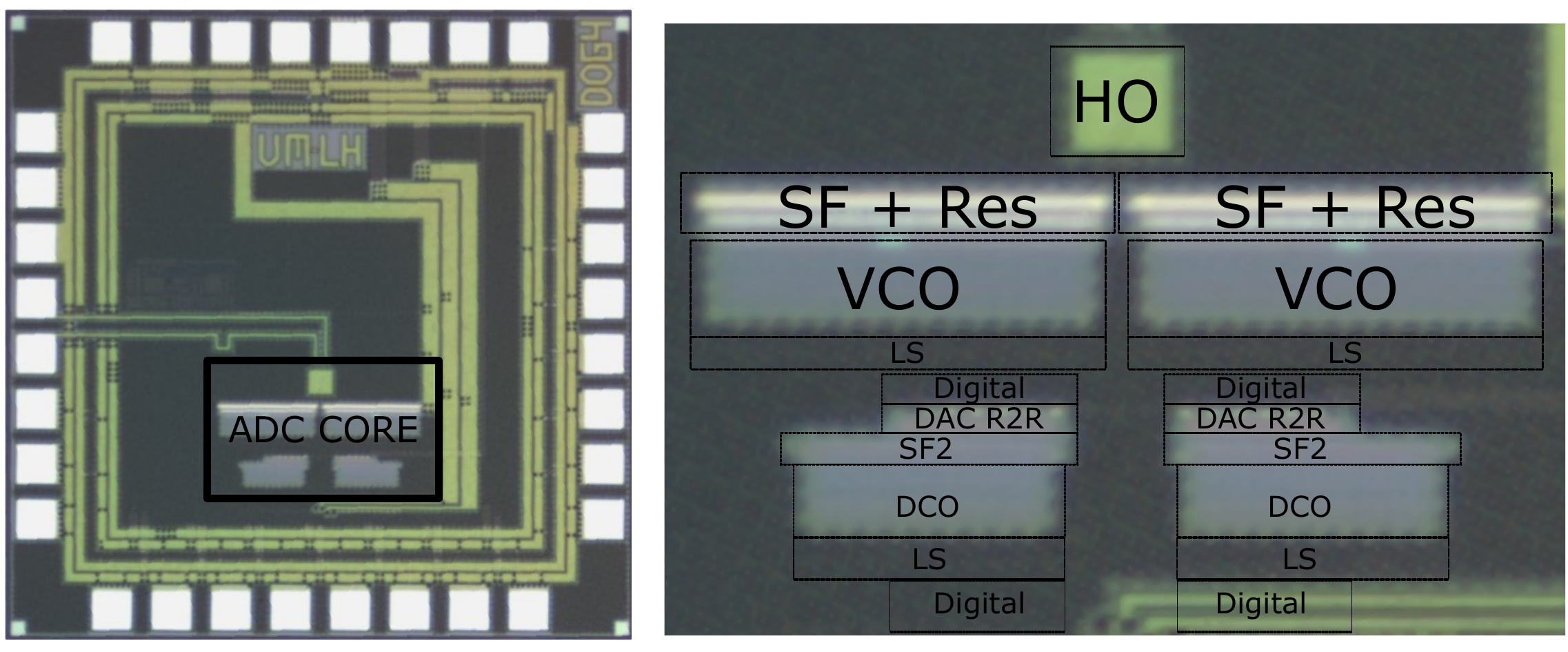}
	\caption{Die micrograph.}
	\label{fig:micrograph}
\end{figure}

The proposed architecture depicted in Fig.~\ref{fig:complete_system} has been used to implement an audio ADC suitable for a MEMS microphone in 130 nm CMOS. The chip has a fully differential analog input AC coupled to an external signal source using a dummy MEMS capacitive coupler implemented in a separate chip. The chip provides a 6 bit parallel output data bus sampled with an external clock of 3.072~MHz. Analog supply pin is connected to 1.5~V. Digital logic is supplied from a separate pin at 0.95~V.  The chip includes also a bandgap reference generator. A micrograph of the fabricated chip is shown in Fig.~\ref{fig:micrograph}(a). The details of the layout of the different circuit blocks can be observed in Fig.~\ref{fig:micrograph}(b). The total area is 0.095~mm2. Almost two-thirds of the chip area is devoted to the first integrator stage due to the VCO and SF size requirements to meet flicker noise specifications. The second integrator and the output stages altogether occupy only one half of the first integrator stage area. Input signals have been generated with a Stanford Instruments DS-360 audio generator and digital data has been captured with a logic analyzer.

The total power consumption is 250~uW, 175~uW correspond to the analog domain and 75~uW to the digital domain. Most of the analog power consumption corresponds to the input integrator stage (SF + Resistor + VCO). According to post-layout simulations, around 10~uW correspond to the bandgap and 12~uW correspond to the second integrator stage. Same as with the area, the power consumption of the second integrator stage plus the output stage are significantly smaller that the first integrator stage. This is due to the more relaxed requirements of the DCO, whose noise and distortion is shaped by the sigma-delta loop.  

\begin{figure}[t]
	\centering
	\includegraphics[width=\columnwidth,keepaspectratio]{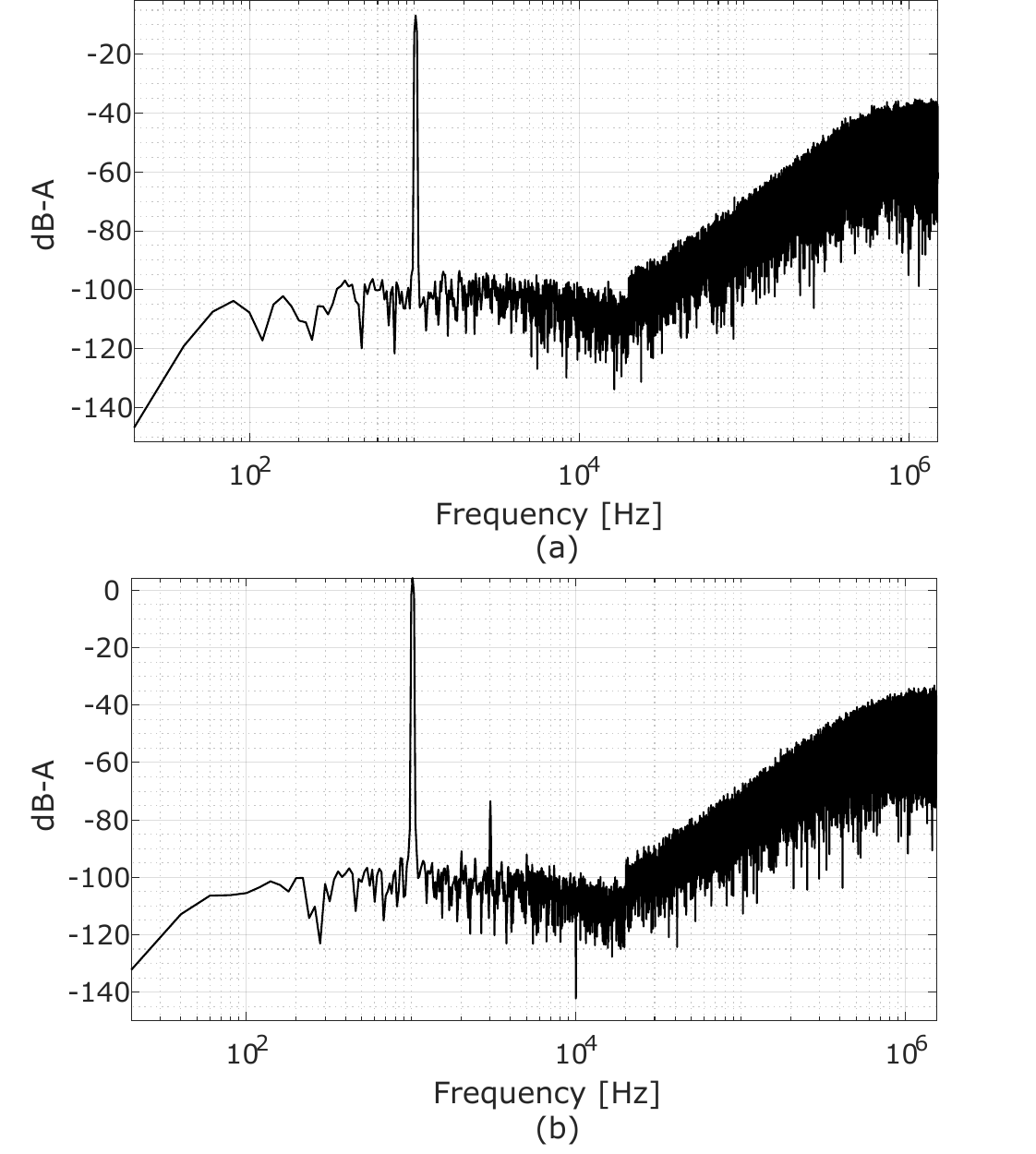}
	\caption{FFT for (a) -36 dBV and (b) -25dBV.}
	\label{fig:fft_-36dBV}
\end{figure}

\begin{figure}[t]
	\centering
	\includegraphics[width=\columnwidth,keepaspectratio]{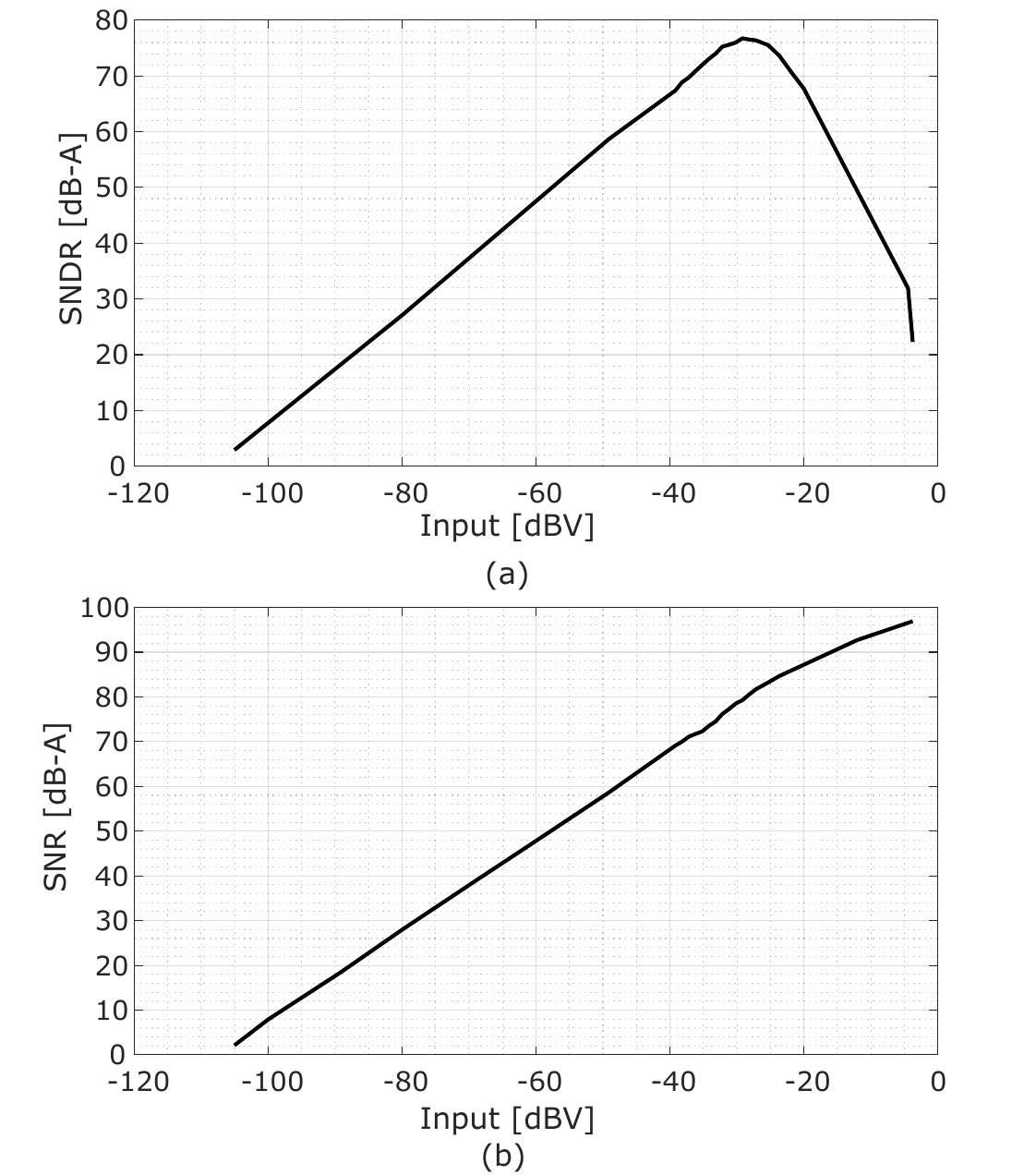}
	\caption{Dynamic range of the measured chip, (a) SNDR, (b) SNR only, distortion components are ignored.}
	\label{fig:measured_DR}
\end{figure}

\begin{figure}[t]
	\centering
	\includegraphics[width=\columnwidth,keepaspectratio]{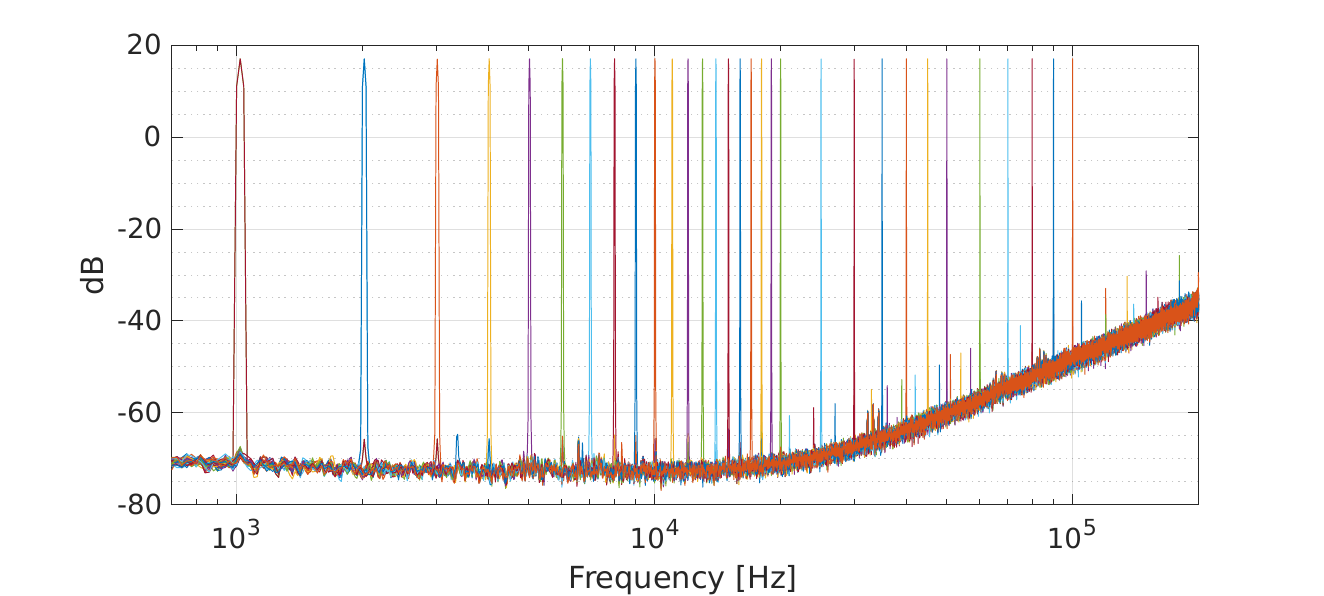}
	\caption{Input frequency sweep (1 kHz to 100 kHz) to evaluate the STF.}
	\label{fig:stf}
\end{figure}

Fig.\ref{fig:fft_-36dBV}(a) shows the output FFT for a -36dBV input signal, which corresponds to the voltage provided by a conventional MEMS under a standard 94~dBSPL audio signal. An A-weighting filter will be used for all measurements, extended only to the audio band (between 20Hz and 20kHz). As can be seen, a quantization noise slope of -40dB/dec is clearly visible, confirming second-order noise-shaping. The SNDR for this input is 69~dB-A. Fig.\ref{fig:fft_-36dBV}(b) shows the same data for a -25~dBV input (40mVp), which shows significant distortion, producing a SNDR of 64.3~dB-A. Fig.~\ref{fig:measured_DR}(a) shows the dynamic rage considering the SNDR, which peaks at a maximum value of 76.5~dB-A. Fig.~\ref{fig:measured_DR}(b) hows the same plot but referred to the Signal-to-noise Ratio (SNR), without distortion. MEMS microphone output is considered valid up to the Acoustic Overload Point (AOP), defined as the amplitude when Total harmonic Distortion (THD) reaches 5\%, \cite{carlos_sensors}, which in our ADC corresponds to -4.4 dBV. According to Fig.~\ref{fig:measured_DR}(b), the dynamic range can be estimated in 103~dB. Fig. \ref{fig:stf} shows the output FFT for an input of -36 dBV and sweeping the frequencies from 1kHz to 100kHz. It can be observed that a third-harmonic component appears for high frequency inputs. This distortion component originates in the second DCO and therefore, appears first-order shaped. In Fig.~\ref{fig:stf} the third harmonic grows with a slope of 20 dB per decade, confirming experimentally this fact. Finally, a 90~dB Power Supply Rejection Ratio (PSRR) has been measured for a 70mV, 1kHz input, comparing the response from the signal port to the analog supply.  

Table \ref{table:comparison} shows a comparison of our chip to other recent audio ADCs in similar CMOS technologies, including zoom \cite{makinwa-audio-adc}, continuous-time sigma-delta \cite{pavan_audio} and VCO-based \cite{mercier_audio, NanSun-ASCC-2022, carlosmls, carlos_sensors}. High input impedance ADCs could be directly driven by capacitive MEMS sensors. Low input impedance ADCs may require a low noise input preamplifier to connect a capacitive MEMS which requires extra power. Therefore, for a fair comparison, we have indicated the kind of input circuitry and impedance in Table I. Column 1 shows the performance parameters of our design, displaying a Schreier FoM of  182~dB (DR) and 155.5~dB (SNDR peak) respectively. Our design is only 2dB below the maximum FoM among the table examples. The size of our ADC is the smallest except for \cite{carlosmls}, even including those in a smaller node. As a further advantage, the input impedance allows the direct connection of a capacitive MEMS without preamplifier, which would compare favourably against the FoM of other ADC topologies \cite{makinwa-audio-adc,pavan_audio}. 

\begin{table*}[t]
  \centering
  \caption{Comparison with Other Works}
  \label{table:comparison}
  \begin{tabular}{|l|c|c|c|c|c|c|c|}
  \hline
   & \textbf{This work}          & \cite{carlos_sensors} & \cite{carlosmls} & \cite{NanSun-ASCC-2022} & \cite{pavan_audio} & \cite{makinwa-audio-adc} & \cite{mercier_audio} \\ \hhline{|=|=|=|=|=|=|=|=|}
   Technology            & \textbf{130 nm} & 130 nm & 130 nm & 28 nm & 180 nm & 160 nm & 65 nm\\ \hline
   Topology                 & VCO-ADC & VCO-ADC & VCO-ADC & VCO-ADC & CTSDM & CT Zoom & VCO-ADC \\ \hline
    Input         & \textbf{Hi-Z}  & Hi-Z & Hi-Z & Hi-Z & Low Z & Low Z & Low Z \\ \hline
   BW [kHz]                 & \textbf{20} & 20 & 20 & 10 & 24 & 24 & 20\\ \hline
   Area [mm2]               & \textbf{0.095} & 0.14 & 0.01 & 0.095 & 0.64 & 0.36 & 0.11 \\ \hline
   Supply (A/D) [V]         & \textbf{1.5 / 0.95} & 1.5 / 0.95 & 1.2 / 0.95 & 1.1 / 0.55 & 1.8 & 1.8 & 1\\ \hline
   Power [uW]               & \textbf{250} & 438.1 & 170.5 & 7.1 & 265 & 590 & 142.7\\ \hline
   Sampling frequency [MHz] & \textbf{3.072} & 3.072 & 2.4 & 0.72 & 6.144 & 5.12 & 2\\ \hline
   DR [dB]                  & \textbf{103} & 108 & 91 & 79.3 & 104 & 108.3 & 100.3\\ \hline
   FOM SNDR [dB]             & \textbf{155.5} & 157 & 152.6 & 170.1 & 180.5 & 182.7 & 175.5\\ \hline
   FOM DR [dB]               & \textbf{182} & 184.6 & 170.5 & 170.8 & 183.6 & 183.4 & 181.8\\ \hline
  \end{tabular}
\end{table*}

\section{Conclusions}
In this paper, we have proven experimentally a practical solution that breaks the dynamic range limitation of prior art high-order True-VCO ADCs. Our design introduces three novelties. Firstly, a new True-VCO architecture is demonstrated by rearranging the building blocks of a conventional second order sigma-delta modulator. Secondly, this new arrangement enables the representation of state variables using modulo arithmetic on finite length binary magnitudes. This is the key contribution to achieve a virtually unlimited dynamic range in the sigma-delta modulator. As a third contribution, we show how the phases of a ring oscillator can be sampled with limited metastability as a Gray code word without requiring the conventional first-difference operation with XOR gates followed by a thermometer-to-binary decoder.  The paper shows a proof-of-concept chip implementation targeted to audio applications where a large dynamic range is a must, while peak SNDR is of lesser concern \cite{carlos_sensors}. The chip has a Schreier FoM of 182~dB (DR), with very small area when compared with other architectures in similar CMOS nodes.  

\section*{Acknowledgement}
We would like to thank Diego Garcia for his help in the measurements.

\bibliography{referencias}

\begin{thebibliography}{10}
\providecommand{\url}[1]{#1}
\csname url@samestyle\endcsname
\providecommand{\newblock}{\relax}
\providecommand{\bibinfo}[2]{#2}
\providecommand{\BIBentrySTDinterwordspacing}{\spaceskip=0pt\relax}
\providecommand{\BIBentryALTinterwordstretchfactor}{4}
\providecommand{\BIBentryALTinterwordspacing}{\spaceskip=\fontdimen2\font plus
\BIBentryALTinterwordstretchfactor\fontdimen3\font minus \fontdimen4\font\relax}
\providecommand{\BIBforeignlanguage}[2]{{%
\expandafter\ifx\csname l@#1\endcsname\relax
\typeout{** WARNING: IEEEtran.bst: No hyphenation pattern has been}%
\typeout{** loaded for the language `#1'. Using the pattern for}%
\typeout{** the default language instead.}%
\else
\language=\csname l@#1\endcsname
\fi
#2}}
\providecommand{\BIBdecl}{\relax}
\BIBdecl

\bibitem{whyandhow}
E.~Gutierrez, P.~Rombouts, and L.~Hernandez, ``{Why and How VCO-based ADCs can improve instrumentation applications},'' in \emph{2018 25th IEEE International Conference on Electronics, Circuits and Systems (ICECS)}, 2018, pp. 101--104.

\bibitem{rvco}
Y.~Lee, B.~Cho, C.~Lee, J.~Kim, and Y.~Chae, ``{A 0.5-ms 47.5-nJ Resistor-to-Digital Converter for Resistive BTEX Sensor Achieving 0.1-to-5 ppb Resolution},'' \emph{IEEE Journal of Solid-State Circuits}, vol.~58, no.~3, pp. 827--837, 2023.

\bibitem{carlosmls}
C.~Perez, A.~Quintero, P.~Amaral, A.~Wiesbauer, and L.~Hernandez, ``{A 73dB-A Audio VCO-ADC Based on a Maximum Length Sequence Generator in 130nm CMOS},'' \emph{IEEE Transactions on Circuits and Systems II: Express Briefs}, vol.~68, no.~10, pp. 3194--3198, 2021.

\bibitem{cardestimestamp}
F.~Cardes, E.~Azizi, and A.~Hierlemann, ``{A Time-Domain Readout Technique for Neural Interfaces Based on VCO-Timestamping},'' \emph{IEEE Transactions on Biomedical Circuits and Systems}, vol.~17, no.~3, pp. 574--584, 2023.

\bibitem{colorines2}
G.~G. Gielen, L.~Hernandez, and P.~Rombouts, ``{Time-Encoding Analog-to-Digital Converters: Bridging the Analog Gap to Advanced Digital CMOS Part 2: Architectures and Circuits},'' \emph{IEEE Solid-State Circuits Magazine}, vol.~12, no.~3, pp. 18--27, 2020.

\bibitem{nan-sun-purely-vco-adc}
Y.~Zhong, S.~Li, X.~Tang, L.~Shen, W.~Zhao, S.~Wu, and N.~Sun, ``{A Second-Order Purely VCO-Based CT $\Delta\Sigma$ ADC Using a Modified DPLL Structure in 40-nm CMOS},'' \emph{IEEE Journal of Solid-State Circuits}, vol.~55, no.~2, pp. 356--368, 2020.

\bibitem{carlos_sensors}
C.~Perez, R.~Garvi, G.~Lopez, A.~Quintero, F.~Leger, P.~Amaral, A.~Wiesbauer, and L.~Hernandez, ``{A VCO-Based ADC With Direct Connection to a Microphone MEMS, 80-dB Peak SNDR and 438-{μW} Power Consumption},'' \emph{IEEE Sensors Journal}, vol.~23, no.~8, pp. 8466--8477, 2023.

\bibitem{nus}
T.-F. Wu and M.~S.-W. Chen, ``{A Noise-Shaped VCO-Based Nonuniform Sampling ADC With Phase-Domain Level Crossing},'' \emph{IEEE Journal of Solid-State Circuits}, vol.~54, no.~3, pp. 623--635, 2019.

\bibitem{galton}
G.~Taylor and I.~Galton, ``{A Reconfigurable Mostly-Digital Delta-Sigma ADC With a Worst-Case FOM of 160 dB},'' \emph{IEEE Journal of Solid-State Circuits}, vol.~48, no.~4, pp. 983--995, 2013.

\bibitem{sensors_andres}
\BIBentryALTinterwordspacing
A.~Quintero, F.~Cardes, C.~Perez, C.~Buffa, A.~Wiesbauer, and L.~Hernandez, ``{A VCO-Based CMOS Readout Circuit for Capacitive MEMS Microphones},'' \emph{Sensors}, vol.~19, no.~19, 2019. [Online]. Available: \url{https://www.mdpi.com/1424-8220/19/19/4126}
\BIBentrySTDinterwordspacing

\bibitem{jaewook_kim_tcasi}
J.~Kim, T.-K. Jang, Y.-G. Yoon, and S.~Cho, ``{Analysis and Design of Voltage-Controlled Oscillator Based Analog-to-Digital Converter},'' \emph{IEEE Transactions on Circuits and Systems I: Regular Papers}, vol.~57, no.~1, pp. 18--30, 2010.

\bibitem{rombouts_electronic_letters}
\BIBentryALTinterwordspacing
A.~Babaie-Fishani and P.~Rombouts, ``{Highly linear VCO for use in VCO-ADCs},'' \emph{Electronics Letters}, vol.~52, no.~4, pp. 268--270, 2016. [Online]. Available: \url{https://ietresearch.onlinelibrary.wiley.com/doi/abs/10.1049/el.2015.3807}
\BIBentrySTDinterwordspacing

\bibitem{garvilin}
R.~Garvi, J.~Granizo, E.~Gutierrez, V.~Medina, A.~Wiesbauer, and L.~Hernandez, ``{A VCO-ADC Linearized by a Capacitive Frequency-to-Current Converter},'' \emph{IEEE Transactions on Circuits and Systems II: Express Briefs}, vol.~70, no.~6, pp. 1841--1845, 2023.

\bibitem{bulkdlr}
J.~Ahmadi-Farsani, V.~Zúñiga-González, T.~Serrano-Gotarredona, B.~Linares-Barranco, and J.~M. de~la Rosa, ``{Enhanced Linearity in FD-SOI CMOS Body-Input Analog Circuits – Application to Voltage-Controlled Ring Oscillators and Frequency-Based ΣΔ ADCs},'' \emph{IEEE Transactions on Circuits and Systems I: Regular Papers}, vol.~67, no.~10, pp. 3297--3308, 2020.

\bibitem{optimization_borgmans}
J.~Borgmans, E.~Sacco, P.~Rombouts, and G.~Gielen, ``{Methodology for Readout and Ring Oscillator Optimization Toward Energy-Efficient VCO-Based ADCs},'' \emph{IEEE Transactions on Circuits and Systems I: Regular Papers}, vol.~69, no.~3, pp. 985--998, 2022.

\bibitem{colorines1}
G.~G. Gielen, L.~Hernandez, and P.~Rombouts, ``{Time-Encoding Analog-to-Digital Converters: Bridging the Analog Gap to Advanced Digital CMOS-Part 1: Basic Principles},'' \emph{IEEE Solid-State Circuits Magazine}, vol.~12, no.~2, pp. 47--55, 2020.

\bibitem{cardes_noise}
\BIBentryALTinterwordspacing
F.~Cardes, A.~Quintero, E.~Gutierrez, C.~Buffa, A.~Wiesbauer, and L.~Hernandez, ``{SNDR Limits of Oscillator-Based Sensor Readout Circuits},'' \emph{Sensors}, vol.~18, no.~2, 2018. [Online]. Available: \url{https://www.mdpi.com/1424-8220/18/2/445}
\BIBentrySTDinterwordspacing

\bibitem{rombouts_true}
\BIBentryALTinterwordspacing
A.~Babaie-Fishani and P.~Rombouts, ``{True high-order VCO-based ADC},'' \emph{Electronics Letters}, vol.~51, no.~1, pp. 23--25, 2015. [Online]. Available: \url{https://ietresearch.onlinelibrary.wiley.com/doi/abs/10.1049/el.2014.2719}
\BIBentrySTDinterwordspacing

\bibitem{cardes-patente-vco}
A.~Wiesbauer, S.~Dietmar, L.~Hernandez, and F.~Cardes, ``{System and method for an oversampled data converter},'' US Patent 2014 0,270,261.

\bibitem{rombouts-third-order}
A.~Babaie-Fishani and P.~Rombouts, ``{A Mostly Digital VCO-Based CT-SDM With Third-Order Noise Shaping},'' \emph{IEEE Journal of Solid-State Circuits}, vol.~52, no.~8, pp. 2141--2153, 2017.

\bibitem{cardes-second-order}
F.~Cardes, E.~Gutierrez, A.~Quintero, C.~Buffa, A.~Wiesbauer, and L.~Hernandez, ``{0.04-mm2 103-dB-A Dynamic Range Second-Order VCO-Based Audio $\Sigma\Delta$ ADC in 0.13- $\mu$m CMOS},'' \emph{IEEE Journal of Solid-State Circuits}, vol.~53, no.~6, pp. 1731--1742, 2018.

\bibitem{vco_integrators}
L.~Hernandez, E.~Gutierrez, and F.~Cardes, ``{Frequency-encoded integrators applied to filtering and sigma-delta modulation},'' in \emph{2016 IEEE International Symposium on Circuits and Systems (ISCAS)}, 2016, pp. 478--481.

\bibitem{quintero_sscl}
A.~Quintero, C.~Buffa, C.~Perez, F.~Cardes, D.~Straeussnigg, A.~Wiesbauer, and L.~Hernandez, ``{A Coarse-Fine VCO-ADC for MEMS Microphones With Sampling Synchronization by Data Scrambling},'' \emph{IEEE Solid-State Circuits Letters}, vol.~3, pp. 29--32, 2020.

\bibitem{pavan_audio}
S.~Billa, S.~Dixit, and S.~Pavan, ``{Analysis and Design of an Audio Continuous-Time 1-X FIR-MASH Delta–Sigma Modulator},'' \emph{IEEE Journal of Solid-State Circuits}, vol.~55, no.~10, pp. 2649--2659, 2020.

\bibitem{mercier_audio}
J.~Huang and P.~P. Mercier, ``{A 94.2-dB SNDR 142.6-μW VCO-Based Audio ADC With a Split-ADC Differential Pulse Code Modulation Architecture},'' \emph{IEEE Solid-State Circuits Letters}, vol.~4, pp. 121--124, 2021.

\bibitem{steyaert_coarse-fine}
J.~Daniels, W.~Dehaene, and M.~Steyaert, ``{All-digital differential VCO-based A/D conversion},'' in \emph{Proceedings of 2010 IEEE International Symposium on Circuits and Systems}, 2010, pp. 1085--1088.

\bibitem{vesterbacka}
V.~Unnikrishnan and M.~Vesterbacka, ``{Time-Mode Analog-to-Digital Conversion Using Standard Cells},'' \emph{IEEE Transactions on Circuits and Systems I: Regular Papers}, vol.~61, no.~12, pp. 3348--3357, 2014.

\bibitem{vco-adc-perrott}
M.~Z. Straayer and M.~H. Perrott, ``{A 12-Bit, 10-MHz Bandwidth, Continuous-Time $\Sigma\Delta$ ADC With a 5-Bit, 950-MS/s VCO-Based Quantizer},'' \emph{IEEE Journal of Solid-State Circuits}, vol.~43, no.~4, pp. 805--814, 2008.

\bibitem{gray_medina}
V.~Medina, R.~Garvi, E.~Gutierrez, S.~Paton, and L.~Hernandez, ``{A Gray-Encoded Ring Oscillator for Efficient Frequency-to-Digital Conversion in VCO-Based ADCs},'' \emph{IEEE Transactions on Circuits and Systems II: Express Briefs}, vol.~70, no.~3, pp. 870--874, 2023.

\bibitem{CASM_CTSDM_PFM}
V.~Medina, P.~Rombouts, and L.~Hernandez, ``{A Different View of Sigma-Delta Modulators Under the Lens of Pulse Frequency Modulation},'' 2023.

\bibitem{Hogenauer}
E.~Hogenauer, ``{An economical class of digital filters for decimation and interpolation},'' \emph{IEEE Transactions on Acoustics, Speech, and Signal Processing}, vol.~29, no.~2, pp. 155--162, 1981.

\bibitem{abidi-noise_CMOS_RO}
A.~Abidi, ``{Phase Noise and Jitter in CMOS Ring Oscillators},'' \emph{IEEE Journal of Solid-State Circuits}, vol.~41, no.~8, pp. 1803--1816, 2006.

\bibitem{borgmans-noise-RO}
J.~Borgmans, R.~Riem, and P.~Rombouts, ``{The Analog Behavior of Pseudo Digital Ring Oscillators Used in VCO ADCs},'' \emph{IEEE Transactions on Circuits and Systems I: Regular Papers}, vol.~68, no.~7, pp. 2827--2840, 2021.

\bibitem{luca-sant_mems_microphone}
L.~Sant, M.~Füldner, E.~Bach, F.~Conzatti, A.~Caspani, R.~Gaggl, A.~Baschirotto, and A.~Wiesbauer, ``{A 130dB SPL 72dB SNR MEMS Microphone Using a Sealed-Dual Membrane Transducer and a Power-Scaling Read-Out ASIC},'' \emph{IEEE Sensors Journal}, vol.~22, no.~8, pp. 7825--7833, 2022.

\bibitem{lanuzza_levelshifters}
M.~Lanuzza, P.~Corsonello, and S.~Perri, ``{Low-Power Level Shifter for Multi-Supply Voltage Designs},'' \emph{IEEE Transactions on Circuits and Systems II: Express Briefs}, vol.~59, no.~12, pp. 922--926, 2012.

\bibitem{medina_analysis_vco_subset_feedback)}
V.~Medina, L.~M. Alvero-Gonzalez, E.~Gutierrez, L.~Hernandez, and S.~Paton, ``{Analysis of VCO-Based Continuous-Time ΣΔ ADCs Using a Subset of Phases as the Feedback Signal},'' \emph{IEEE Transactions on Circuits and Systems II: Express Briefs}, pp. 1--1, 2024.

\bibitem{makinwa-audio-adc}
S.~Mehrotra, E.~Eland, S.~Karmakar, A.~Liu, B.~Gönen, M.~Bolatkale, R.~Van~Veldhoven, and K.~A. Makinwa, ``{A 590 µW, 106.6 dB SNDR, 24 kHz BW Continuous-Time Zoom ADC with a Noise-Shaping 4-bit SAR ADC},'' in \emph{ESSCIRC 2022- IEEE 48th European Solid State Circuits Conference (ESSCIRC)}, 2022, pp. 253--256.

\bibitem{NanSun-ASCC-2022}
Y.~Zhong, L.~Jie, and N.~Sun, ``{A 78.6 dB-SNDR 520mVpp-full-scale 620MΩ-Zin 105dBCMRR VCO-based Sensor Readout Circuit Using FVF-Based Gm-Input Structure},'' in \emph{2022 IEEE Asian Solid-State Circuits Conference (A-SSCC)}, 2022, pp. 1--3.

\end{thebibliography}
\bibliographystyle{IEEEtran}
\vspace{11pt}

\begin{IEEEbiography}[{\includegraphics[width=1in,height=1.25in,clip,keepaspectratio]{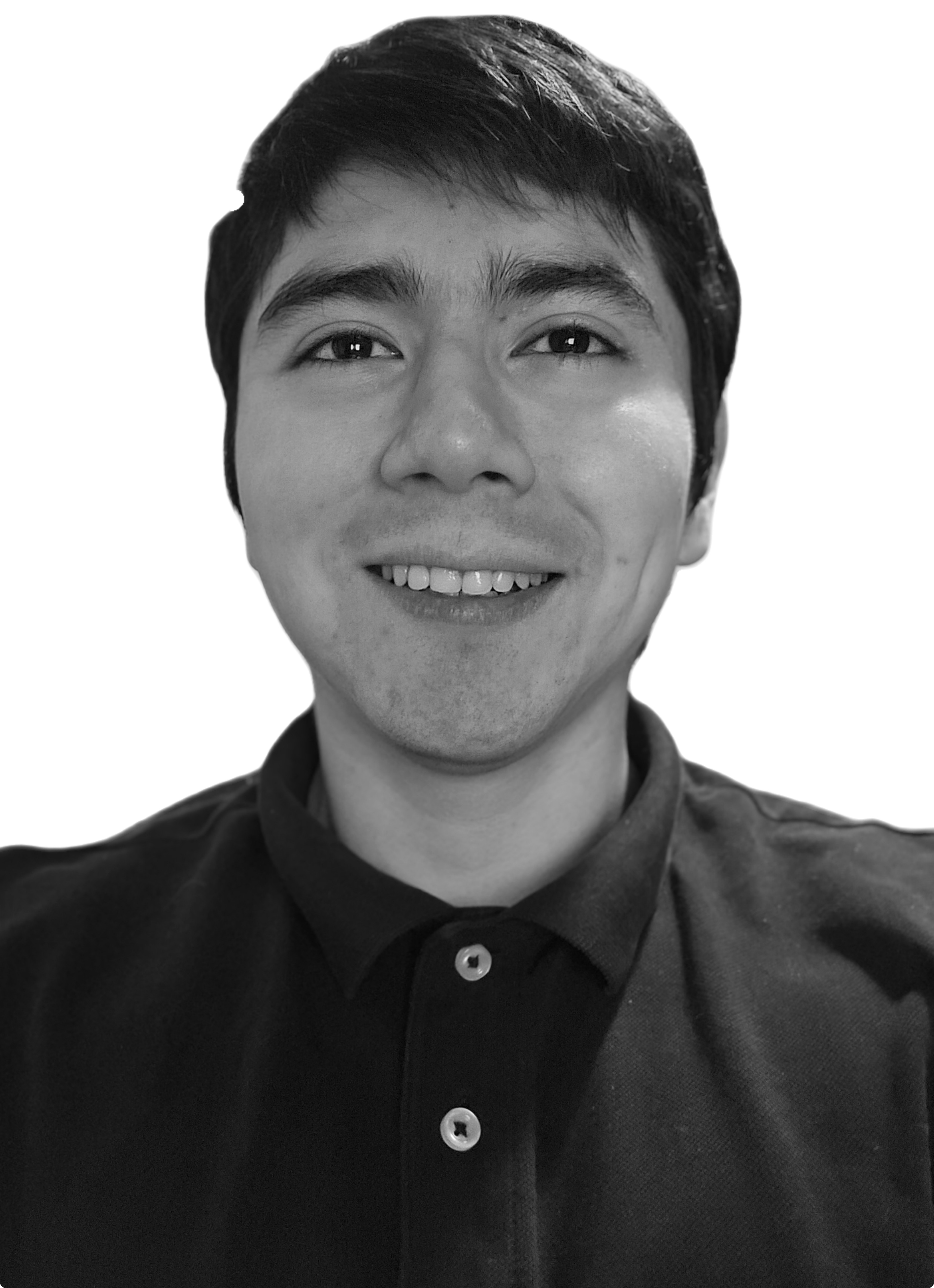}}]{Victor Medina} received the B.Sc. and M.Sc. degrees in Electronic Engineering from Carlos III University of Madrid, Spain, in 2018 and 2019, where he is currently pursuing the Ph.D. degree. In 2022, he did a four-month internship at Infineon Technologies, Villach, Austria. His current research interests include mixed-signal integrated circuit design and time-encoded systems theory.
\end{IEEEbiography}

\begin{IEEEbiography}[{\includegraphics[width=1in,height=1.25in,clip,keepaspectratio]{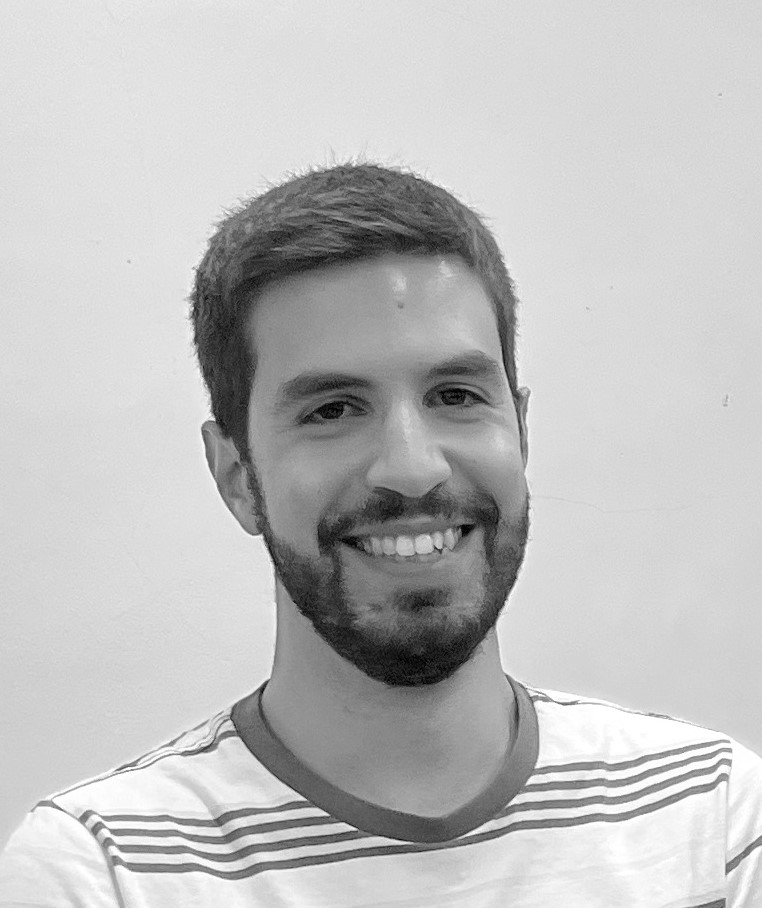}}]{Ruben Garvi Jimenez-Ortiz} received the B.Sc., M.E. and PhD degrees in Electronic engineering from Carlos III University, Madrid, Spain, in 2016, 2017 and 2023 respectively. In 2019 he did a four month internship at Infineon Technologies Austria. His current research interests include mixed-signal integrated circuits design and MEMS sensors. He is currently an Assistant Professor at Carlos III University.
\end{IEEEbiography}

\begin{IEEEbiography}[{\includegraphics[width=1in,height=1.25in,clip,keepaspectratio]{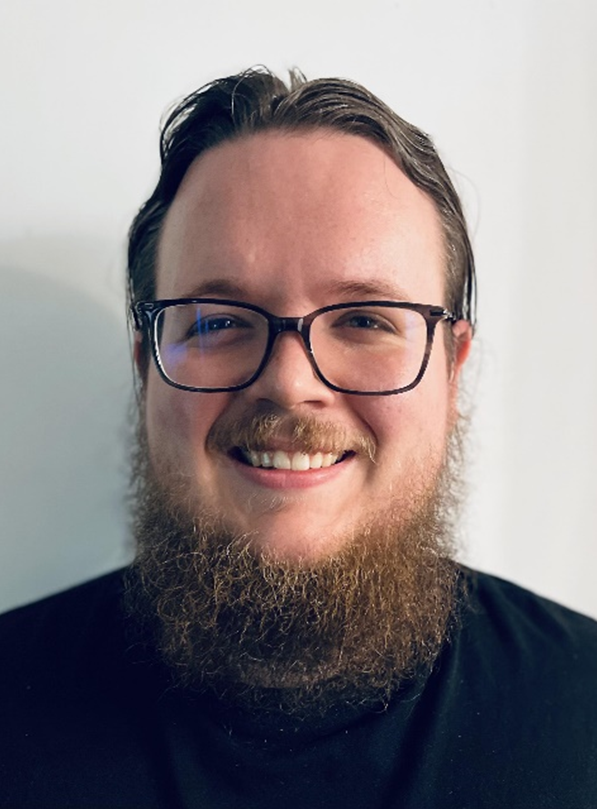}}]{Javier Granizo} received the M.E. degree in electronic engineering from Carlos III University of Madrid in 2021, where he is currently pursuing a Ph.D. degree. 
His current research interests include oscillator-based ADCs, mixed-signal circuits, and Asynchronous Neural Network CMOS accelerators. 
\end{IEEEbiography}

\begin{IEEEbiography}[{\includegraphics[width=1in,height=1.25in,clip,keepaspectratio]{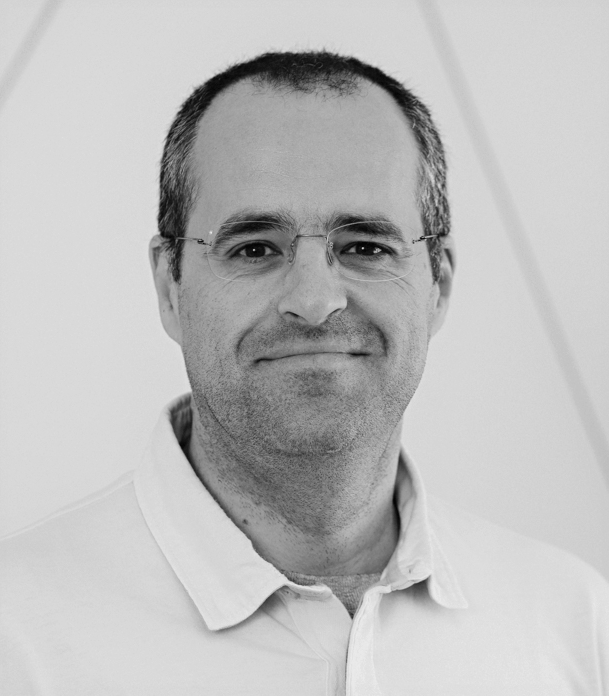}}]{Pedro Amaral}
 received the master’s degree in applied physics from the University of Lisbon, Lisbon, Portugal, in 1997, and the Ph.D. degree in experimental particle physics from the University of Chicago, Chicago, IL, USA, 2003.
From 2005 to 2007, he was a Postdoctoral Research Fellow at CERN (Genéve), Meyrin, Switzerland. From 2007 to 2012, he worked on PLL and RF design for Synopsys. Since 2012, he has been an Analog/Concept Engineer at Infineon, Villach, Austria, focusing on PLL and ADC design for automotive applications, and Silicon Microphone design. He has authored four papers in experimental particle physics and medical physics. He has coauthored over 120 papers in experimental particle physics. He has three patents granted in PLL, microphone and ultrasonic sensors, and three other patents pending.
\end{IEEEbiography}

\begin{IEEEbiography}[{\includegraphics[width=1in,height=1.25in,clip,keepaspectratio]{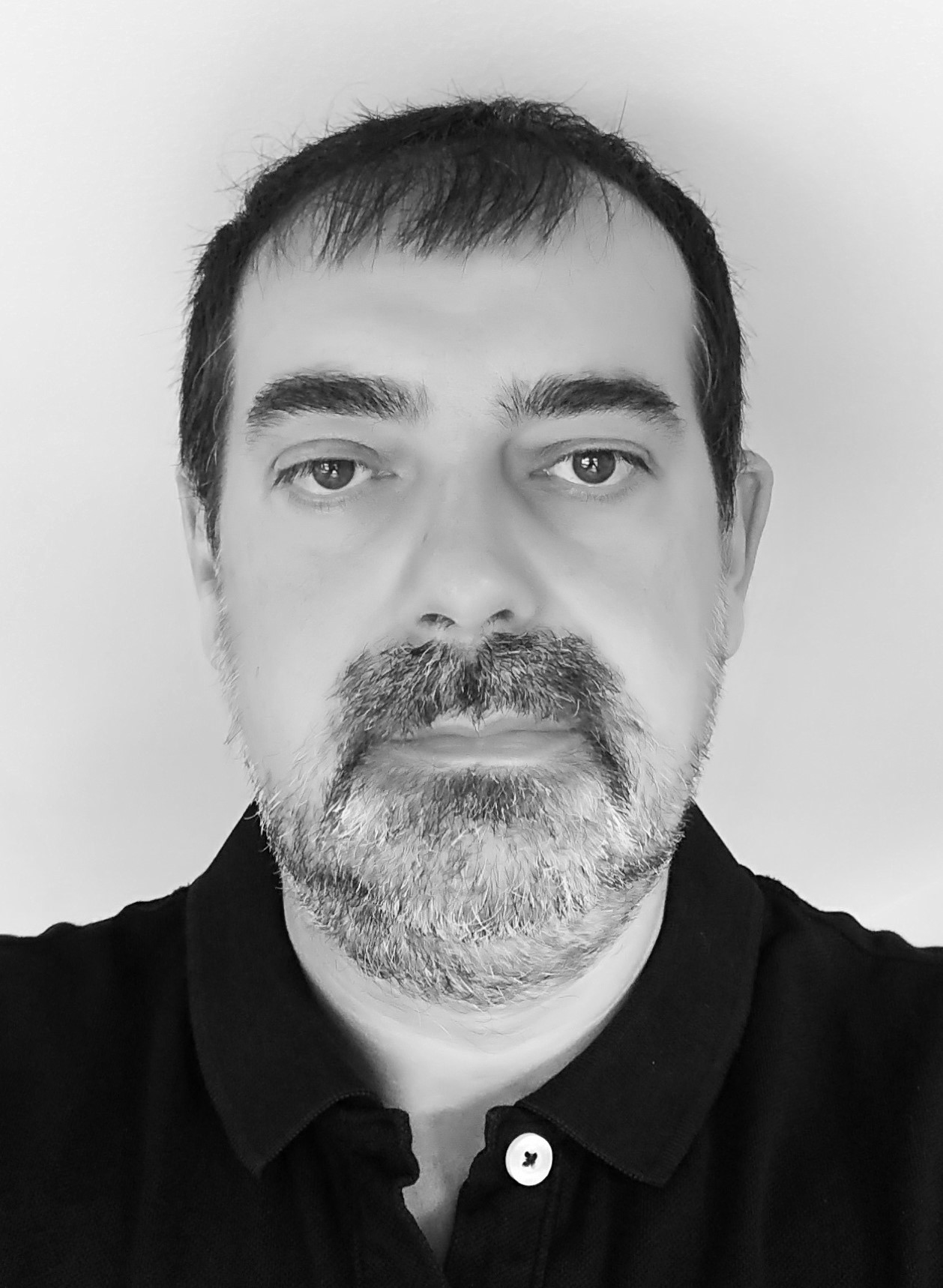}}]{Luis Hernandez-Corporales} received a MS (‘89) and PhD (’95) degrees in Telecommunication Engineering from Polytechnic University of Madrid. He did a postdoctoral stay at the ECE dept. of Oregon State University in 1996 and Analog Devices in Willmington, USA in 1997. In 1998 he joined University Carlos III of Madrid where he is currently Full Professor in the Electronic Technology Department and leads the mixed signal research group. He has been department head and PhD program director. In 2009 he did a sabbatical stay at IMEC, Leuven, Belgium. He has coauthored three books, over 170 papers and holds 25 patents. He is a member of the IEEE-CAS ASPTC committee and has been associate editor of IEEE Transactions on Circuits and systems I and II for 9 years. His topics of interest are analog microelectronics, Sigma-Delta modulation, time-encoded data converters and neural networks.
\end{IEEEbiography}

\vfill

\end{document}